\documentclass[tighten,times,twocolumn]{aastex62}
\usepackage{apjfonts}
\usepackage{psfig}
\usepackage{float}

\usepackage{graphics}
\usepackage{amssymb,amsmath}
\usepackage{color}

\newcommand{\degree}{\ensuremath{^\circ}}

\newcommand{\chisq}{\ensuremath{\chi^2}}
\newcommand{\msun}{\ensuremath{\rm{M}_\odot}}
\newcommand{\rsun}{\ensuremath{\rm{R}_\odot}}

\newcommand{\swift}{{\it Swift}}

\newcommand{\sneia}{SNe Ia}
\newcommand{\snia}{SN Ia}

\newcommand{\tAbt}{\ensuremath{ 2458144.850^{+ 0.001 }_{- 0.001 } }} 
\newcommand{\tBbt}{\ensuremath{ 4.373^{+ 0.020 }_{- 0.016 } }} 
\newcommand{\aAbt}{\ensuremath{ 1.167^{+ 0.004 }_{- 0.003 } }} 
\newcommand{\aBbt}{\ensuremath{ 1.393^{+ 0.005 }_{- 0.007 } }}

\newcommand{\triseKbt}{\ensuremath{ 18.125^{+ 0.008 }_{- 0.008 } }} 
\newcommand{\triseBbt}{\ensuremath{ 18.150^{+ 0.297 }_{- 0.297 } }} 
 
\newcommand{\tdetbt}{\ensuremath{ 1.438^{+ 0.027 }_{- 0.000 } }}

\newcommand{\tAKSNb}{\ensuremath{ 2455827.6^{+ 0.6 }_{- 0.3 } }} 
\newcommand{\tBKSNb}{\ensuremath{ 2.6^{+ 0.6 }_{- 1.1 } }} 
\newcommand{\aAKSNb}{\ensuremath{ 1.9^{+ 0.7 }_{- 0.4 } }} 
\newcommand{\aBKSNb}{\ensuremath{ 1.9^{+ 0.2 }_{- 0.1 } }}

\newcommand{\triseKKSNb}{\ensuremath{ 18.7^{+ 0.4 }_{- 0.6 } }} 
\newcommand{\triseBKSNb}{\ensuremath{ 18.3^{+ 0.6 }_{- 0.8 } }} 
 
\newcommand{\tdetKSNb}{\ensuremath{ 55^{+ 17 }_{- 6 } }}

\newcommand{\tAKSNc}{\ensuremath{ 2455907.4^{+ 2.1 }_{- 1.2 } }} 
\newcommand{\tBKSNc}{\nodata } 
\newcommand{\aAKSNc}{\ensuremath{ 2.2^{+ 1.1 }_{- 0.4 } }} 
\newcommand{\aBKSNc}{\nodata }

\newcommand{\triseKKSNc}{\ensuremath{ 19.1^{+ 1.1 }_{- 1.8 } }} 
\newcommand{\triseBKSNc}{\ensuremath{ 18.8^{+ 1.1 }_{- 1.8 } }} 
 
\newcommand{\tdetKSNc}{\ensuremath{ 131^{+ 71 }_{- 6 } }}

\newcommand{\tAKSNa}{\ensuremath{ 2456161.1^{+ 0.2 }_{- 0.2 } }} 
\newcommand{\tBKSNa}{\ensuremath{ 4.3^{+ 0.3 }_{- 0.3 } }} 
\newcommand{\aAKSNa}{\ensuremath{ 1.4^{+ 0.1 }_{- 0.1 } }} 
\newcommand{\aBKSNa}{\ensuremath{ 1.07^{+ 0.05 }_{- 0.08 } }}

\newcommand{\triseKKSNa}{\ensuremath{ 15.1^{+ 0.3 }_{- 0.3 } }} 
\newcommand{\triseBKSNa}{\ensuremath{ 14.8^{+ 0.5 }_{- 0.5 } }} 
 
\newcommand{\tdetKSNa}{\ensuremath{ 22^{+ 20 }_{- 2 } }}

\begin{document}

\title{Seeing Double: ASASSN-18bt Exhibits a Two-Component Rise in the Early-Time {\em K2}  Light Curve}  
\shorttitle{ASASSN-18bt} 
\shortauthors{Shappee et al. }

\author[0000-0003-4631-1149]{B.~J.~Shappee}
\affiliation{Institute for Astronomy, University of Hawai'i, 2680 Woodlawn Drive, Honolulu, HI 96822, USA}

\author[0000-0001-9206-3460]{T.~W.-S.~Holoien}
\altaffiliation{Carnegie Fellow}
\affiliation{The Observatories of the Carnegie Institution for Science, 813 Santa Barbara St., Pasadena, CA 91101, USA}

\author{M.~R.~Drout}
\altaffiliation{Hubble Fellow, Dunlap Fellow}
\affiliation{The Observatories of the Carnegie Institution for Science, 813 Santa Barbara St., Pasadena, CA 91101, USA}

\author{K.~Auchettl}
\affiliation{Center for Cosmology and AstroParticle Physics (CCAPP), The Ohio State University, 191 W.\ Woodruff Ave., Columbus, OH 43210, USA}
\affiliation{Department of Physics, The Ohio State University, 191 W. Woodruff Avenue, Columbus, OH 43210, USA}
\affiliation{Dark Cosmology Centre, Niels Bohr Institute, University of Copenhagen, Blegdamsvej 17, 2100 Copenhagen, Denmark}

\author{M.~D.~Stritzinger}
\affiliation{Department of Physics and Astronomy, Aarhus University, Ny Munkegade 120, DK-8000 Aarhus C, Denmark}
\affiliation{Visiting Astronomer, Institute for Astronomy, University of Hawai’i, 2680 Woodlawn Drive, Honolulu, HI 96822, USA}

\author{C.~S.~Kochanek}
\affiliation{Center for Cosmology and AstroParticle Physics (CCAPP), The Ohio State University, 191 W.\ Woodruff Ave., Columbus, OH 43210, USA}
\affiliation{Department of Astronomy, The Ohio State University, 140 West 18th Avenue, Columbus, OH 43210, USA}

\author{K.~Z.~Stanek}
\affiliation{Center for Cosmology and AstroParticle Physics (CCAPP), The Ohio State University, 191 W.\ Woodruff Ave., Columbus, OH 43210, USA}
\affiliation{Department of Astronomy, The Ohio State University, 140 West 18th Avenue, Columbus, OH 43210, USA}

\author{E.~Shaya}
\affiliation{Astronomy Department, University of Maryland, College Park, MD 20742-2421, USA.}

\author{G.~Narayan}
\affiliation{Department of Physics and Astronomy, Johns Hopkins University, Baltimore, MD 21218, USA.}

\collaboration{}
\collaboration{\textit{ASAS-SN}}

\author{J.~S.~Brown}
\affiliation{Department of Astronomy, The Ohio State University, 140 West 18th Avenue, Columbus, OH 43210, USA}

\author{S.~Bose}
\affiliation{Kavli Institute for Astronomy and Astrophysics, Peking University, Yi He Yuan Road 5, Hai Dian District, Beijing 100871, China}

\author{D.~Bersier}
\affiliation{Astrophysics Research Institute, Liverpool John Moores University, 146 Brownlow Hill, Liverpool L3 5RF, UK}

\author{J.~Brimacombe}
\affiliation{Coral Towers Observatory, Cairns, Queensland 4870, Australia}

\author{Ping~Chen}
\affiliation{Kavli Institute for Astronomy and Astrophysics, Peking University, Yi He Yuan Road 5, Hai Dian District, Beijing 100871, China}

\author{Subo~Dong}
\affiliation{Kavli Institute for Astronomy and Astrophysics, Peking University, Yi He Yuan Road 5, Hai Dian District, Beijing 100871, China}

\author{S.~Holmbo}
\affiliation{Department of Physics and Astronomy, Aarhus University, Ny Munkegade 120, DK-8000 Aarhus C, Denmark}

\author{B.~Katz}
\affiliation{Department of Particle Physics and Astrophysics, Weizmann Institute of Science, Rehovot 76100, Israel}

\author{J.~A.~Mu\~{n}oz}
\affiliation{Departamento de Astronom\'{\i}a y Astrof\'{\i}sica, Universidad de Valencia, E-46100 Burjassot, Valencia, Spain}
\affiliation{Observatorio Astron\'omico, Universidad de Valencia, E-46980 Paterna, Valencia, Spain}

\author{R.~L.~Mutel}
\affiliation{Department of Physics and Astronomy, University of Iowa, Iowa City, IA 52242, USA}

\author{R.~S.~Post}
\affiliation{Post Observatory, Lexington, MA 02421, USA}

\author{J.~L.~Prieto}
\affiliation{N\'ucleo de Astronom\'ia de la Facultad de Ingenier\'ia y Ciencias, Universidad Diego Portales, Av. Ej\'ercito 441, Santiago, Chile}
\affiliation{Millennium Institute of Astrophysics, Santiago, Chile}

\author{J.~Shields}
\affiliation{Department of Astronomy, The Ohio State University, 140 West 18th Avenue, Columbus, OH 43210, USA}

\author{D.~Tallon}
\affiliation{Department of Physics and Astronomy, University of Iowa, Iowa City, IA 52242, USA}

\author{T.~A.~Thompson}
\affiliation{Center for Cosmology and AstroParticle Physics (CCAPP), The Ohio State University, 191 W.\ Woodruff Ave., Columbus, OH 43210, USA}
\affiliation{Department of Astronomy, The Ohio State University, 140 West 18th Avenue, Columbus, OH 43210, USA}

\author{P.~J.~Vallely}
\affiliation{Department of Astronomy, The Ohio State University, 140 West 18th Avenue, Columbus, OH 43210, USA}

\author{S.~Villanueva~Jr.}
\affiliation{Department of Astronomy, The Ohio State University, 140 West 18th Av., Columbus, OH 43210, USA}

\collaboration{}
\collaboration{\textit{ATLAS}}

\author{L.~Denneau}
\affiliation{Institute of Astronomy, University of Hawai'i, 2680 Woodlawn Drive, Honolulu, HI 96822, USA.}

\author{H.~Flewelling}
\affiliation{Institute of Astronomy, University of Hawai'i, 2680 Woodlawn Drive, Honolulu, HI 96822, USA.}

\author{A.~N.~Heinze}
\affiliation{Institute of Astronomy, University of Hawai'i, 2680 Woodlawn Drive, Honolulu, HI 96822, USA.}

\author{K.~W.~Smith}
\affiliation{Astrophysics Research Centre, School of Mathematics and Physics, Queens University Belfast, Belfast BT7 1NN, UK.}

\author{B.~Stalder}
\affiliation{LSST, 950 North Cherry Avenue, Tucson, AZ 85719.}

\author{J.~L.~Tonry}
\affiliation{Institute of Astronomy, University of Hawai'i, 2680 Woodlawn Drive, Honolulu, HI 96822, USA.}

\author{H.~Weiland}
\affiliation{Institute of Astronomy, University of Hawai'i, 2680 Woodlawn Drive, Honolulu, HI 96822, USA.}

\collaboration{}
\collaboration{\textit{Kepler/K2}}

\author{T.~Barclay}
\affiliation{NASA Goddard Space Flight Center, 8800 Greenbelt Rd, Greenbelt, MD 20771, USA.}
\affiliation{University of Maryland, Baltimore County, 1000 Hilltop Cir, Baltimore, MD 21250, USA.}

\author{G.~Barentsen}
\affiliation{NASA Ames Research Center, Moffett Field, CA 94035, USA.}
\affiliation{Bay Area Environmental Research Institute, P.O. Box 25, Moffett Field, CA 94035, USA.}

\author{A.~M.~Cody}
\affiliation{NASA Ames Research Center, Moffett Field, CA 94035, USA.}
\affiliation{Bay Area Environmental Research Institute, P.O. Box 25, Moffett Field, CA 94035, USA.}

\author{J.~Dotson}
\affiliation{NASA Ames Research Center, Moffett Field, CA 94035, USA.}

\author{F.~Foerster}
\affiliation{Center for Mathematical Modeling, University of Chile, Santiago, Chile.}

\author{P.~Garnavich}
\affiliation{Department of Physics, University of Notre Dame, 225 Nieuwland Science Hall, Notre Dame, IN, 46556-5670, USA.}

\author{M.~Gully-Santiago}
\affiliation{NASA Ames Research Center, Moffett Field, CA 94035, USA.}
\affiliation{Bay Area Environmental Research Institute, P.O. Box 25, Moffett Field, CA 94035, USA.}

\author{C.~Hedges}
\affiliation{NASA Ames Research Center, Moffett Field, CA 94035, USA.}
\affiliation{Bay Area Environmental Research Institute, P.O. Box 25, Moffett Field, CA 94035, USA.}

\author{S.~Howell}
\affiliation{NASA Ames Research Center, Moffett Field, CA 94035, USA.}

\author{D.~Kasen}
\affiliation{Department of Astronomy, University of California, Berkeley, CA 94720-3411, USA.}
\affiliation{Lawrence Berkeley National Laboratory, 1 Cyclotron Road, Berkeley, California 94720, USA.}

\author{S.~Margheim}
\affiliation{Gemini Observatory, La Serena, Chile.}

\author{R.~Mushotzky}
\affiliation{Astronomy Department, University of Maryland, College Park, MD 20742-2421, USA.}

\author{A.~Rest}
\affiliation{Space Telescope Science Institute, 3700 San Martin Drive, Baltimore, MD 21218, USA.}
\affiliation{Department of Physics and Astronomy, Johns Hopkins University, Baltimore, MD 21218, USA.}

\author{B.~E.~Tucker}
\affiliation{The Research School of Astronomy and Astrophysics, Mount Stromlo Observatory, Australian National University, via Cotter Road, Canberra, ACT 2611, Australia.}
\affiliation{The ARC Centre of Excellence for All-Sky Astrophysics (CAASTRO).}
\affiliation{National Centre for the Public Awareness of Science, Australian National University, Canberra, ACT 2601, Australia}

\author{A.~Villar}
\affiliation{Harvard-Smithsonian Center for Astrophysics, 60 Garden Street, Cambridge, MA 02138, USA.}

\author{A.~Zenteno}
\affiliation{Cerro Tololo Inter-American Observatory, Casilla 603, La Serena, Chile.}

\collaboration{}
\collaboration{\textit{Kepler Spacecraft Team}}

\author{G.~Beerman}
\affiliation{LASP, University of Colorado at Boulder, Boulder, CO 80303, USA}

\author{R.~Bjella}
\affiliation{LASP, University of Colorado at Boulder, Boulder, CO 80303, USA}

\author{G.~Castillo}
\affiliation{LASP, University of Colorado at Boulder, Boulder, CO 80303, USA}

\author{J.~Coughlin}
\affiliation{NASA Ames Research Center, Moffett Field, CA 94035, USA.}
\affiliation{SETI Institute, 189 Bernardo Avenue, Mountain View, CA 94043, USA}

\author{B.~Elsaesser}
\affiliation{LASP, University of Colorado at Boulder, Boulder, CO 80303, USA}

\author{S.~Flynn}
\affiliation{LASP, University of Colorado at Boulder, Boulder, CO 80303, USA}

\author{R.~Gangopadhyay}
\affiliation{LASP, University of Colorado at Boulder, Boulder, CO 80303, USA}

\author{K.~Griest}
\affiliation{LASP, University of Colorado at Boulder, Boulder, CO 80303, USA}

\author{M.~Hanley}
\affiliation{LASP, University of Colorado at Boulder, Boulder, CO 80303, USA}

\author{J.~Kampmeier}
\affiliation{LASP, University of Colorado at Boulder, Boulder, CO 80303, USA}

\author{R.~Kloetzel}
\affiliation{LASP, University of Colorado at Boulder, Boulder, CO 80303, USA}

\author{L.~Kohnert}
\affiliation{LASP, University of Colorado at Boulder, Boulder, CO 80303, USA}

\author{C.~Labonde}
\affiliation{LASP, University of Colorado at Boulder, Boulder, CO 80303, USA}

\author{R.~Larsen}
\affiliation{LASP, University of Colorado at Boulder, Boulder, CO 80303, USA}

\author{K.~A.~Larson}
\affiliation{Ball Aerospace and Technologies Corp., Boulder, Colorado, 80301, USA}

\author{K.~M.~McCalmont-Everton}
\affiliation{Ball Aerospace and Technologies Corp., Boulder, Colorado, 80301, USA}

\author{C.~McGinn}
\affiliation{LASP, University of Colorado at Boulder, Boulder, CO 80303, USA}

\author{L.~Migliorini}
\affiliation{LASP, University of Colorado at Boulder, Boulder, CO 80303, USA}

\author{J.~Moffatt}
\affiliation{LASP, University of Colorado at Boulder, Boulder, CO 80303, USA}

\author{M.~Muszynski}
\affiliation{LASP, University of Colorado at Boulder, Boulder, CO 80303, USA}

\author{V.~Nystrom}
\affiliation{LASP, University of Colorado at Boulder, Boulder, CO 80303, USA}

\author{D.~Osborne}
\affiliation{LASP, University of Colorado at Boulder, Boulder, CO 80303, USA}

\author{M.~Packard}
\affiliation{LASP, University of Colorado at Boulder, Boulder, CO 80303, USA}

\author{C.~A.~Peterson}
\affiliation{Ball Aerospace and Technologies Corp., Boulder, Colorado, 80301, USA}

\author{M.~Redick}
\affiliation{LASP, University of Colorado at Boulder, Boulder, CO 80303, USA}

\author{L.~H.~Reedy}
\affiliation{LASP, University of Colorado at Boulder, Boulder, CO 80303, USA}

\author{S.~E.~Ross}
\affiliation{Ball Aerospace and Technologies Corp., Boulder, Colorado, 80301, USA}

\author{B.~Spencer}
\affiliation{LASP, University of Colorado at Boulder, Boulder, CO 80303, USA}

\author{K.~Steward}
\affiliation{LASP, University of Colorado at Boulder, Boulder, CO 80303, USA}

\author{J.~E.~Van Cleve}
\affiliation{NASA Ames Research Center, Moffett Field, CA 94035, USA.}
\affiliation{SETI Institute, 189 Bernardo Avenue, Mountain View, CA 94043, USA}

\author{J.~Vin\'icius de Miranda\ Cardoso}
\affiliation{NASA Ames Research Center, Moffett Field, CA 94035, USA.}
\affiliation{Universidade Federal de Campina Grande, Campina Grande, Brazil}

\author{T.~Weschler}
\affiliation{LASP, University of Colorado at Boulder, Boulder, CO 80303, USA}

\author{A.~Wheaton}
\affiliation{LASP, University of Colorado at Boulder, Boulder, CO 80303, USA}

\collaboration{}
\collaboration{\textit{Pan-STARRS}}

\author{J.~Bulger}
\affiliation{Institute of Astronomy, University of Hawai'i, 2680 Woodlawn Drive, Honolulu, HI 96822, USA.}

\author{K.~C.~Chambers}
\affiliation{Institute of Astronomy, University of Hawai'i, 2680 Woodlawn Drive, Honolulu, HI 96822, USA.}

\author{H.~A.~Flewelling}
\affiliation{Institute of Astronomy, University of Hawai'i, 2680 Woodlawn Drive, Honolulu, HI 96822, USA.}

\author{M.~E.~Huber}
\affiliation{Institute of Astronomy, University of Hawai'i, 2680 Woodlawn Drive, Honolulu, HI 96822, USA.}

\author{T.~B.~Lowe}
\affiliation{Institute of Astronomy, University of Hawai'i, 2680 Woodlawn Drive, Honolulu, HI 96822, USA.}

\author{E.~A.~Magnier}
\affiliation{Institute of Astronomy, University of Hawai'i, 2680 Woodlawn Drive, Honolulu, HI 96822, USA.}

\author{A.~S.~B.~Schultz}
\affiliation{Institute of Astronomy, University of Hawai'i, 2680 Woodlawn Drive, Honolulu, HI 96822, USA.}

\author{C.~Z.~Waters}
\affiliation{Institute of Astronomy, University of Hawai'i, 2680 Woodlawn Drive, Honolulu, HI 96822, USA.}

\author{M.~Willman}
\affiliation{Institute of Astronomy, University of Hawai'i, 2680 Woodlawn Drive, Honolulu, HI 96822, USA.}

\collaboration{}
\collaboration{\textit{PTSS / TNTS}}

\author{E.~Baron}
\affiliation{Homer L. Dodge Department of Physics and Astronomy, University of Oklahoma, Norman, OK}
\affiliation{Visiting Astronomer at the Department of Physics and Astronomy, Aarhus University, Ny Munkegade 120, DK-8000 Aarhus C, Denmark}

\author{Zhihao Chen}
\affiliation{Physics Department and Tsinghua Center for Astrophysics (THCA), Tsinghua University, Beijing, 100084, China}

\author{James M. Derkacy}
\affiliation{Homer L. Dodge Department of Physics and Astronomy, University of Oklahoma, Norman, OK}

\author{Fang Huang}
\affiliation{Department of Astronomy, School of Physics and Astronomy, Shanghai Jiao Tong University, Shanghai, 200240, China}
\affiliation{Physics Department and Tsinghua Center for Astrophysics (THCA), Tsinghua University, Beijing, 100084, China}

\author{Linyi Li}
\affiliation{Physics Department and Tsinghua Center for Astrophysics (THCA), Tsinghua University, Beijing, 100084, China}

\author{Wenxiong~Li}
\affiliation{Physics Department and Tsinghua Center for Astrophysics (THCA), Tsinghua University, Beijing, 100084, China}
\affiliation{Las Cumbres Observatory, 6740 Cortona Drive, Suite 102, Goleta, CA 93117-5575, USA}

\author{Xue Li}
\affiliation{Physics Department and Tsinghua Center for Astrophysics (THCA), Tsinghua University, Beijing, 100084, China}

\author{Jun~Mo}
\affil{Physics Department and Tsinghua Center for Astrophysics (THCA), Tsinghua University, Beijing, 100084, China}

\author{Liming Rui}
\affiliation{Physics Department and Tsinghua Center for Astrophysics (THCA), Tsinghua University, Beijing, 100084, China}

\author{Hanna Sai}
\affiliation{Physics Department and Tsinghua Center for Astrophysics (THCA), Tsinghua University, Beijing, 100084, China}

\author{Lifan Wang}
\affiliation{Purple Mountain Observatory, Chinese Academy of Sciences, Nanjing 210034, China}
\affiliation{George P. and CynthiaWoods Mitchell Institute for Fundamental Physics $\&$ Astronomy, Texas A. $\&$ M.
University, Department of Physics and Astronomy, 4242 TAMU, College Station, TX 77843, USA}

\author{Lingzhi Wang}
\affiliation{National Astronomical Observatory of China, Chinese Academy of Sciences, Beijing, 100012, China}

\author{Xiaofeng~Wang}
\affiliation{Physics Department and Tsinghua Center for Astrophysics (THCA), Tsinghua University, Beijing, 100084, China}

\author{Danfeng Xiang}
\affiliation{Physics Department and Tsinghua Center for Astrophysics (THCA), Tsinghua University, Beijing, 100084, China}

\author{Jicheng Zhang}
\affiliation{Physics Department and Tsinghua Center for Astrophysics (THCA), Tsinghua University, Beijing, 100084, China}

\author{Jujia Zhang}
\affiliation{Yunnan Astronomical Observatory of China, Chinese Academy of Sciences, Kunming, 650011, China}

\author{Kaicheng Zhang}
\affiliation{Physics Department and Tsinghua Center for Astrophysics (THCA), Tsinghua University, Beijing, 100084, China}
\affiliation{Department of Astronomy, University of Texas at Austin, Austin, TX, 78712, USA}

\author{Tianmeng Zhang}
\affiliation{National Astronomical Observatory of China, Chinese Academy of Sciences, Beijing, 100012, China}

\author{Xinghan Zhang}
\affiliation{Physics Department and Tsinghua Center for Astrophysics (THCA), Tsinghua University, Beijing, 100084, China}

\author{Xulin Zhao}
\affiliation{School of Science, Tianjin University of Technology, Tianjin, 300384, China}

\collaboration{}
\nocollaboration

\author{P.~J.~Brown}
\affiliation{ Department of Physics and Astronomy,
Texas A\&M University, 4242 TAMU, College Station, TX 77843, USA}
\affiliation{ George P. and Cynthia Woods Mitchell Institute for Fundamental Physics and Astronomy}

\author{J.~J.~Hermes}
\altaffiliation{Hubble Fellow}
\affiliation{Department of Physics and Astronomy, University of North Carolina,
Chapel Hill, NC 27599, USA}

\author{J.~Nordin}
\affiliation{Institute of Physics, Humboldt-Universit\"at zu Berlin, Newtonstr. 15, 12489 Berlin, Germany}

\author{S.~Points}
\affiliation{Cerro Tololo Inter-American Observatory, Casilla 603, La Serena, Chile.}

\author{A.~S\'odor}
\affiliation{Konkoly Observatory, MTA CSFK, Konkoly Thege M. ut 15-17, Budapest, 1121, Hungary}

\author{G.~M. Strampelli}
\affiliation{Space Telescope Science Institute, 3700 San Martin Drive, Baltimore, MD 21218, USA.}
\affiliation{University of La Laguna,  Calle Padre Herrera, 38200 San Cristóbal de La Laguna, Santa Cruz de Tenerife, Spain.}

\author{A.~Zenteno}
\affiliation{Cerro Tololo Inter-American Observatory, Casilla 603, La Serena, Chile.}

\correspondingauthor{B.~J.~Shappee}
\email{shappee@hawaii.edu}

\date{\today}

\begin{abstract}

On 2018 Feb.~4.41, the All-Sky Automated Survey for SuperNovae (ASAS-SN) discovered ASASSN-18bt in the {\em K2}  Campaign 16 field. With a redshift of $z=0.01098$ and a peak apparent magnitude of $B_{max}=14.31$,  ASASSN-18bt is the nearest and brightest SNe Ia yet observed by the {\em Kepler}  spacecraft.  Here we present the discovery of ASASSN-18bt, the {\em K2}  light curve, and pre-discovery data from ASAS-SN and the Asteroid Terrestrial-impact Last Alert System (ATLAS). The {\em K2}  early-time light curve has an unprecedented 30-minute cadence and photometric precision for an SN~Ia light curve, and it unambiguously shows a $\sim 4$ day nearly linear phase followed by a steeper rise. Thus, ASASSN-18bt joins a growing list of SNe Ia whose early light curves are not well described by a single power law. We show that a double-power-law model fits the data reasonably well, hinting that two physical processes must be responsible for the observed rise. However, we find that current models of the interaction with a non-degenerate companion predict an abrupt rise and cannot adequately explain the initial, slower linear phase.  Instead, we find that existing, published models with shallow $^{56}$Ni are able to span the observed behavior and, with tuning, may be able to reproduce the ASASSN-18bt light curve. Regardless, more theoretical work is needed to satisfactorily model this and other early-time SNe~Ia light curves.  Finally, we use \swift{} X-ray non-detections to constrain the presence of circumstellar material (CSM) at much larger distances and lower densities than possible with the optical light curve. For a constant density CSM these non-detections constrain $\rho<\rm{4.5 \times 10^{5}\,\rm{cm^{-3}}}$ at a radius of 4 $\times$ 10$^{15}$ cm from the progenitor star.  Assuming a wind-like environment, we place mass-loss limits of $\dot M<\rm{8 \times 10^{-6}}\,\rm{M_{\sun}yr^{-1}}$ for $v_\textrm{w}=100\,\rm{km\,s^{-1}}$, ruling out some symbiotic progenitor systems.
This work highlights the power of well-sampled early-time data and the need for immediate, multi-band, high-cadence followup for progress in understanding SNe~Ia.

\end{abstract}

\section{Introduction}
\label{sec:introduc}

Type Ia supernovae (\sneia{}) are widely thought to result from the thermonuclear explosion of a carbon-oxygen white dwarf (WD; \citealp{hoyle60}) in a close binary system. However, the exact physical nature of the progenitor systems of \sneia{} is not known, and two competing classes of models remain. In the single-degenerate (SD) model, the WD accretes material from a non-degenerate companion, eventually triggering a thermonuclear runaway \citep{whelan73, nomoto82}. In the double-degenerate (DD) model, the companion is another WD, and a runaway reaction is triggered by the merger  of the two WDs, caused either by the removal of energy and angular momentum through gravitational radiation \citep[e.g.,][]{tutukov79, iben84, webbink84}, or by the perturbations of a third \citep[e.g.,][]{thompson11, katz12, shappee13c, antognini14} or fourth \citep{pejcha13, fang18} body. Searches for observational features that could distinguish between these models have proven difficult, as current simulations based on both the SD \citep[e.g.,][]{kasen09} and  DD violent merger models \citep[e.g.,][]{pakmor12} provide equally accurate models for the observations of \sneia{} around $B$-band maximum light ($t_{B {\rm max}}$).

Several observational tests for the SD model arise from the fact that the companion is struck by the ejecta from the supernova shortly after explosion. First, interaction between the ejecta and the companion modifies the early rise of the light curve. The observational consequences depend on the viewing angle, with the strongest effect occurring when the companion is along the line of sight between the observer and the SN. At a fixed viewing angle, emission from this shock interaction scales proportionally with the radius of the companion $R_\textrm{c}$, and this allows early-time observations to constrain the properties of the companion \citep{kasen10}. Another observational signature comes from the stripping of material from the companion when it is struck by ejecta from the supernova \citep[e.g.,][]{wheeler75, marietta00}. Hydrodynamic simulations from \citet{pan12} and \citet{liu12} showed that approximately $0.1 \-- 0.2$ \msun{} of solar-metallicity material is expected to be removed from a main-sequence (MS) companion. Lastly, the interaction between the ejecta and the companion is also expected to affect the future properties of the companion \citep[e.g.,][]{podsiadlowski03, pan12b, shappee13b}. Together, these highlight the need for detailed observational studies of \sneia{} at very early and late times to search for these signatures.

In the past decade, almost two dozen SNe~Ia have been discovered early and have relatively well-sampled early-time light curves.  Surprisingly, \citet{stritzinger18} recently showed that there are two distinct populations of early-time behaviors.  One population exhibits blue colors that slowly evolve and the other population shows red colors and evolves more rapidly. The rising part of SN~Ia light curves also show interesting diversity.  Empirically, the early light curves of some SNe~Ia are reasonably well-fit by a single power law function (e.g., \citealp{nugent11, bloom12, goobar15}) and others show a $2-4$ day nearly-linear rise and then an exponential rise (e.g., \citealp{contreras18}). Finally, many of these well-observed SNe placed limits on masses/radii of a possible companion. These include SN~2009ig ($< 6$ \msun{}; \citealp{foley12}), SN~2011fe ($< 0.1 - 0.25$ \rsun{}; \citealp{bloom12, goobar15}), KSN~2011a ($< 2$ \msun{}; \citealp{olling15}), KSN~2011b ($< 2$ \msun{}; \citealp{olling15}), SN~2012cg ($< 0.24$ \rsun{}; \citealt{silverman12, marion16, shappee18}), SN~2012fr \citep{contreras18}, SN~2013dy ($< 0.35$ \rsun{}; \citealp{zheng13}), SN~2013gy ($< 4$ \rsun{}; \citealp{holmbo18}), SN~2014J ($\lesssim 0.25 - 4$ \rsun{}; \citealp{goobar15, siverd15}), ASASSN-14lp ($\lesssim 0.34 - 11$ \rsun{}; \citealp{shappee16a}), SN~2015F ($< 1.0$ \rsun{}; \citealp{im15, cartier17}), iPTF16abc \citep{miller18}, MUSSES1604D \citep{jiang17}, and DLT~17u (SN~2017cbv; \citealp{hosseinzadeh17}).

The {\em Kepler}  spacecraft has also obtained a number of early-time SN light curves \citep[e.g.,][]{olling15,garnavich16}. Though SNe detected by {\em Kepler}  are rare compared to the numbers found by dedicated transient surveys, {\em Kepler}  light curves can be especially illuminating due to the high, 30-minute cadence and photometric stability of the observations. Previously, 3 SNe~Ia have been observed by {\em Kepler} , providing some of the best early light curve sampling available to date, and none of these light curves show signs of interactions with a stellar companion \citep{olling15}. 

Here we announce the discovery of the Type~Ia SN ASASSN-18bt (SN~2018oh) in UGC 04780 which was monitored by the {\em K2}  mission and analyze the early evolution of the exquisite {\em K2}  light curve. With a peak apparent magnitude of $B_{max}=14.31\pm0.03$ (Li et al. 2018) and a distance of 47.7~Mpc, it is nearer and brighter than any other supernova detected by {\em Kepler} . In Section~\ref{sec:disc_obs}, we describe our discovery and observations of ASASSN-18bt. In Section~\ref{sec:K2_fit}, we analyze the {\em K2}  light curve and find that it is best-fit with a double-power-law model, implying that there may be two different timescales important for describing the rise of ASASSN-18bt. In Section~\ref{sec:companions}, we find that the emission in the first few days seen in the {\em K2}  light curve cannot be described using only models of the interaction with a SD companion. In Section~\ref{sec:NiDist}, we find that the rising light curve also cannot be adequately described using published models that smoothly vary the radioactive $^{56}$Ni distribution in the ejecta, although these models do span the observed behavior of the ASASSN-18bt light curve. In Section~\ref{sec:circumstellar}, we also find that the early-time light curves are also inconsistent with published models for interactions with nearby circumstellar material. In Section~\ref{sec:xray}, X-ray observations are used to place a limit on the mass loss rates prior to explosion. Finally, a summary of our results and a discussion of the implications for the progenitor system and explosion properties of ASASSN-18bt are presented in Section~\ref{sec:summary}.

This work is part of a number of papers analyzing ASASSN-18bt, with coordinated papers from Dimitriadis et al. (2018) and Li et al. (2018). Li et al. (2018) investigates the near-max optical properties of ASASSN-18bt and find $\Delta_{m15}=0.96\pm0.03$ mag, $B_{max}=14.31\pm0.03$ mag, $V_{max}=14.37\pm0.03$ mag, $E(B-V)_{MW}=0.04$, $E(B-V)_{host}=0\pm0.04$ mag, and $t_{Bmax}=58162.7\pm0.3$ day.  Li et al. (2018) also find that the light curve of ASASSN-18bt is consistent with the MW reddening inferred from dust maps alone with no additional host-galaxy reddening.  This is supported by the lack of observed Na ID absorption at the host galaxy's recession velocity.  Using Fit~6 in Table~9 of \citet{folatelli10} and the properties derived from the supernova light curve we estimate the distance to UGC 04780 to be d=$49\pm3$ Mpc.  This distance is consistent with the redshift (47.7 Mpc for $z=0.01098$ and $H_0=69.6$, $\Omega_M=0.286$, $\Omega_\Lambda= 0.714$; \citealp{schneider90}) and is used throughout this work. 

\section{Discovery and Observations}
\label{sec:disc_obs}

The All-Sky Automated Survey for SuperNovae (ASAS-SN; \citealp{shappee14}) is an ongoing project to monitor the entire visible sky with rapid cadence with the aim to discover bright and nearby transients with an unbiased search method. To do this, we use units of four 14-cm lenses on a common mount hosted by the Las Cumbres Observatory global telescope network \citep{brown13} at multiple sites around the globe. After expanding our network in 2017, we currently have five units located in Hawaii, Chile, Texas, and South Africa, allowing us to observe the entire sky every $\sim20$ hours, weather permitting, to a depth of $g\simeq18.5$ mag.  As part of the community effort to support {\em K2}  campaign 16 \citep{howell14,borucki16}, ASAS-SN was monitoring the {\em K2}  field with an increased cadence.  The effort of monitoring the entire {\em K2}  field-of-view multiple times per day was continued through Campaign 17 and will be extended to monitor the {\em TESS}  fields $4-6$ times per day. 

ASASSN-18bt was discovered at J2000 coordinates of $\textrm{RA}=09^{\rm h}06^{\rm m}39.\!\!^{\rm{s}}54$  $\textrm{Dec}=+19\degree20'17.\!\!''77$ in $V$-band images obtained by the ASAS-SN unit ``Brutus'', located on Haleakala, in Hawaii on 2018-02-04.410 UT and was promptly announced to the community \citep{brown18ATel}. 
The {\em K2}  field was monitored by all five ASAS-SN units but, unfortunately, ASASSN-18bt exploded while we were still building reference images on the three recently deployed units and it was only discovered when a post-explosion image was obtained using an older unit.    Worse, the field was not observed between 2018-01-29 and 2018-02-03 because of the fields proximity (within $\sim30$ degrees) to the moon. If it were not for these factors ASASSN-18bt would have been discovered substantially earlier. 
Within 6.8 hours after the discovery, the Asteroid Terrestrial-impact Last Alert System (ATLAS; \citealp{tonry18}) confirmed the source. Almost simultaneously, \citet{leadbeater18TNS} spectroscopically classified ASASSN-18bt as an SN~Ia based on an $R\sim150$ spectrum obtained using the modified ALPY spectrograph at Three Hills Observatory\footnote{As described here \url{http://www.threehillsobservatory.co.uk/astro/spectroscopy.htm}.}. Finally, in \citet{cornect18ATel} we gave an improved position of ASASSN-18bt and presented additional photometry obtained by one of our recently deployed ASAS-SN $g$-band units. The analysis of the {\em K2}  light curve had to wait until the end of Campaign 16, 2018-02-25, when the data was downloaded from the {\em Kepler}  spacecraft and became available. 

\begin{figure*}
\begin{minipage}{\textwidth}
\centering
{\includegraphics[width=0.95\textwidth]{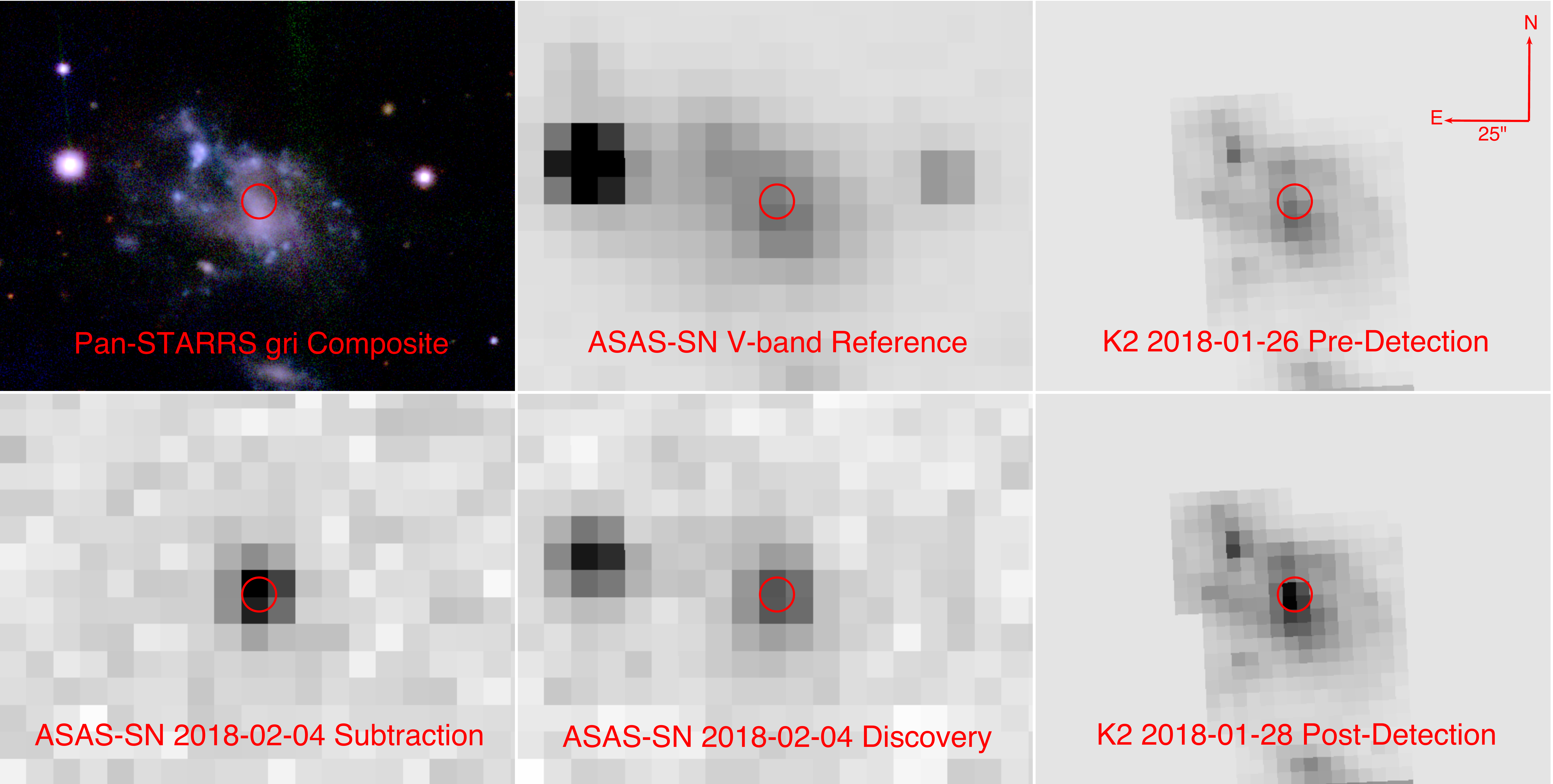}}
\caption{Pre- and post-discovery images of ASASSN-18bt and its host galaxy from Pan-STARRS, ASAS-SN, and {\em K2} . The top-left panel shows a color composite of $g$-, $r$-, and $i$-band images of the host from Pan-STARRS; the top-middle panel shows the ASAS-SN $V$-band reference image of the host; and the top-right panel shows a {\em K2}  pre-detection image obtained on 2018-01-26. The bottom-left panel shows the ASAS-SN $V$-band subtraction image from the epoch of discovery, the bottom-middle panel shows the ASAS-SN $V$-band discovery image, and the bottom-right panel shows a {\em K2}  image from 2018-01-28, after the supernova is visible. The red circle in each image has a radius of 5\farcs{0} and is centered on the position of the SN. A compass and scale are given in the top-right panel for reference.}
\label{fig:discovery}
\end{minipage}
\end{figure*}

Figure~\ref{fig:discovery} shows the reference image, 2018-02-04 discovery image, and the 2018-02-04 first detection difference image from the ASAS-SN ba camera in the top-middle, bottom-middle, and bottom-left panels of the figure, respectively. The 2018-01-26 pre-detection and 2018-01-28 post-detection images of the supernova and its host from {\em K2}  are shown in the top-right and bottom-right panels of the figure, and the top-left panel shows a $gri$-band composite color image of the host galaxy constructed with images obtained by Panoramic Survey Telescope \& Rapid Response System (Pan-STARRS; \citealp{chambers16, flewelling16}). The discovery difference image from ASAS-SN shows that the supernova is clearly detected and the host flux and flux from nearby stars is cleanly subtracted.

The host galaxy of ASASSN-18bt is UGC~04780 ($z=0.01098$, \citealp{schneider90}), a blue barred spiral galaxy with blue clumps in its arms, indicating the likely presence of ongoing star formation. Using archival photometry from Pan-STARRS (optical), the Galaxy Evolution Explorer (GALEX; ultraviolet), and the Wide-field Infrared Explorer (WISE; near-infrared), we fit the spectral energy distribution of UGC 04780 with the publicly available Fitting and Assessment of Synthetic Templates \citep[\textsc{fast};~][]{kriek09}. Given the clumpy nature of the light distribution, we measure the optical magnitudes from the PS1 images by hand and find $g\sim14.9$ mag, $r\sim14.5$ mag, $i\sim14.5$ mag, $z\sim14.4$ mag, and $y\sim14.3$ mag. We assumed a \citet{cardelli89} extinction law with $R_V=3.1$ and a Galactic extinction of $A_V = 0.124$ mag \citep{schlafly11} and employed an exponentially declining star-formation history, a Salpeter initial mass function, and the \citet{bruzual03} stellar population models. Based on the FAST fit, the host galaxy has a stellar mass of $(4.68^{+0.33}_{-0.61})\times10^8$~\msun{} and a star formation rate of $\lesssim 0.05$~\msun~yr$^{-1}$, which is largely consistent with the results from the MPA-JHU Galspec pipeline. However, the galaxy light is dominated by a young stellar population, and the modeling has difficulty fitting both the optical and infrared data, suggesting that our mass estimate should be regarded as an upper limit.

\subsection{ASAS-SN light curve}
\label{sec:ASASSN_LC}

\begin{figure*}
\begin{minipage}{\textwidth}
\centering
{\includegraphics[width=0.95\textwidth]{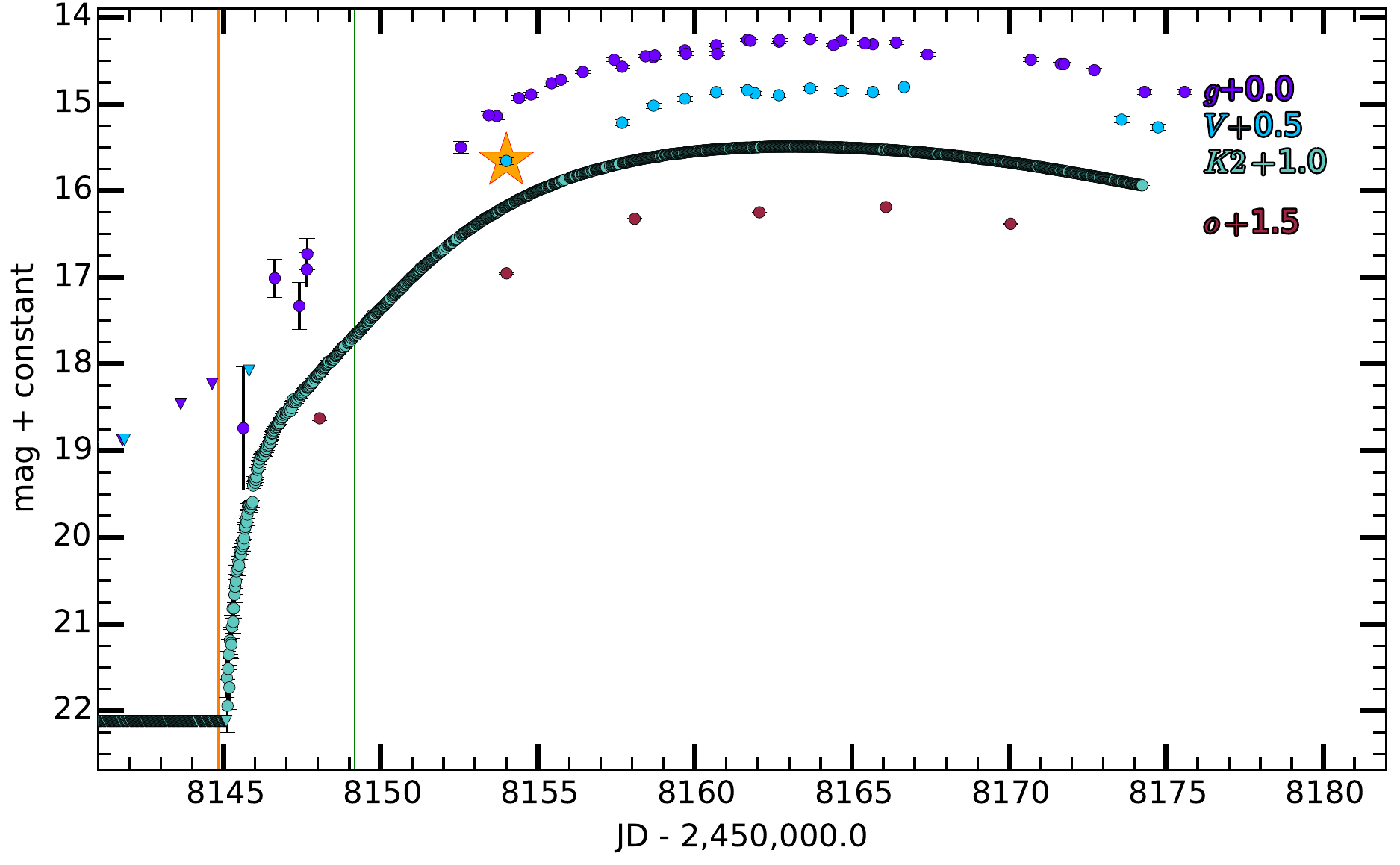}}
\caption{Host-subtracted light curves of ASASSN-18bt from ASAS-SN ($V$ and $g$ filters), {\em K2}  ({\em Kepler}  filter), and ATLAS ('orange' or $o$ filter). 3-sigma limits are shows as downward triangles for epochs where the supernova was not detected. The orange and green vertical bars indicate $t_{\rm{1}}$ and $t_{\rm{2}}$, respectively, as determined by a double-power-law fit to the {\em K2}  light curve (see Section~\ref{sec:K2_fit} and Equation~\ref{eq:double}). The orange star highlights the ASAS-SN discovery epoch of ASASSN-18bt.}
\label{fig:phot}
\end{minipage}
\end{figure*}

ASAS-SN images are processed by the fully automatic ASAS-SN pipeline using the ISIS image subtraction package \citep{alard98, alard00}. A host-galaxy reference image was constructed for each of the ASAS-SN units using images obtained prior to the discovery of ASASSN-18bt, and these were used to subtract the host's background in all science images. Science images that were obviously affected by clouds were removed.  We then performed aperture photometry with a 2-pixel ($\approx 16.\!\!''0$) aperture on each host-template subtracted science image using the IRAF {\tt apphot} package. Photometry of the supernova was calibrated relative to a number of stars in the field of the host galaxy with known magnitudes from the AAVSO Photometric All-Sky Survey (APASS; \citealp{henden15}). The ASAS-SN detections and 3-sigma limits are presented in Table \ref{tab:phot} and plotted in Figure~\ref{fig:phot}.  Throughout the paper, light curves are plotted in observed time and measured rise times are translated to the rest frame.

\subsection{{\em K2}  light curve}
\label{sec:K2_LC}

The {\em K2}  mission is a follow-up to the highly successful {\em Kepler} mission. {\em K2} was instigated when a second reaction wheel was lost, leaving the spacecraft with only two wheels rather than the three required for full 3D stabilization. The best solution for mitigating this problem was to constrain the spacecraft to point in the ecliptic plane, balancing solar wind pressure about the center of mass and minimizing the torques on the spacecraft that rotate the field around the line of sight axis. Thrusters are used every few hours to return the pointing back to a starting orientation, resulting in a sawtooth motion in the positions of stars that is typically on the order of one pixel. This sawtooth pattern is reflected in the photometric counts, but can be reduced by summing over more pixels in a larger aperture, at the cost of introducing more photon noise and contamination from neighboring sources. {\em K2}  also has long-term (weeks and months) sensitivity trends partly due to temperature changes as the Sun angle and zodiacal light levels change within a campaign. {\em Kepler} and {\em K2} have a broad response function from $\sim$420--900~nm \citep{koch10}.

When {\em K2}  Campaign 16 ended, all data for the campaign were downloaded from the spacecraft. The unique nature of the {\em K2}  mission requires careful reduction.  Unfortunately, the relevant CCD channel had moving bands of an electronic pattern called the rolling bands during the observation.  This is a not uncommon occurrence on {\em K2} , and there are flags in the quality arrays that indicate when it passes over the optimal aperture for a target. Because the pattern is fairly constant along a row, we were able to minimize its effects by subtracting the mean at the edges of the downloaded target pixel map (after ignoring pixels which appear to have galaxy or star light).  From examining the other galaxies in the channel with this problem we find that this noise is usually reduced to a level below the shot noise of the background light.  The data taken when the rolling bands were present in the ASASSN-18bt aperture were mostly constrained within 3 days of $t_{\rm{1}}$ (as fit in Section \ref{sec:K2_fit}).  To remove the sawtooth pattern created by changes in the amount of light overfilling the aperture, as {\em K2}  nods due to solar wind pressure,  third order polynomials in two dimensions of centroidal motion were fit to all galaxies observed on the same channel, except for those clearly undergoing variability.  To remove longer time scale trends, we obtain basis vectors from a PCA analysis of these LCs. The LCs on this channel can then be approximated as linear superposition of these vectors plus a unique sawtooth pattern for each galaxy. However, the solutions for the sawtooth patterns remains poor as long as the trending vectors are poor and vice versa. Therefore, an iterative scheme is applied in which we put the long term trends back into the LCs, rerun the PCA analysis and solve for improved basis vectors.  Then after solving for the coefficients of both the sawtooth fit and the trending vectors again, we repeat the procedure. After about a dozen iterations, the procedure converges for the most common 5 trending vectors.

The coefficients to apply to the trending vectors are found by minimizing the variation of the LC after dividing by the linear superposition of the PCA vectors.  This works well because most galaxies have constant brightness over the campaign.  But for a galaxy with a transient like ASASSN-18bt, we are confined to using only the part of the LC with quiet time before and/or after the event. Fortunately, the optimal number of PCA vectors for ASASSN-18bt was just two, and there was a long period in the campaign before eruption to use to determine their coefficients well.

An additional complication is created because a SN moves the center of light from the center of the galaxy towards the SN.  This induces a slight change to the sawtooth function.  Therefore, after solving for the best sawtooth and long term instrumental trending during the quiet time, the sawtooth pattern is removed from the time when the SN exceeds 50\% of the galaxy contribution and a new sawtooth pattern is obtained.  This time the trending is assumed to be valid and the goodness of fit is a measure of how well the corrected LC fits the pattern after smoothing over three or four nodding periods.

Finally, we calibrated the {\em K2}  light curve using the mangled SED from fitting the PS r-band (presented in Li et al. 2018) around peak to determine the synthetic $K2$ peak magnitude and the absolute zeropoint to the {\em K2}  light curve. 
The {\em K2}  detections and 3-sigma limits are shown in Figure~\ref{fig:phot} and, for completeness, presented in Table \ref{tab:phot}.

\begin{deluxetable}{lrcc}
\tablewidth{210pt}
\tabletypesize{\footnotesize}
\tablecaption{Photometric Observations}
\tablehead{
\colhead{JD} &
\colhead{Band} &
\colhead{Magnitude} &
\colhead{Telescope} \\ 
\colhead{($-$2,450,000)} &
\colhead{} &
\colhead{}  &
\colhead{}  }
\startdata
8105.761 & $g$ & >18.69 & ASAS-SN/bi \\ 
7908.470 & $V$ & >17.57 & ASAS-SN/be \\ 
8095.490 & $K2$ & >21.12 & K2 \\ 
8148.053 & $o$ & 17.126(0.028) & ATLAS \\ 
\enddata \tablecomments{$V\--\textrm{band}$ photometry is calibrated in the Vega magnitude system. The Kepler and SDSS $g$-band photometry are calibrated in the AB magnitude system. \textit{Only the first observation in each band is shown here to demonstrate its form and content. Table is included in its entirety as an ancillary file.}} 
\label{tab:phot} 
\end{deluxetable}

\subsection{ATLAS light curve}
\label{sec:ATLAS_LC}

ATLAS is an ongoing survey project primarily designed to detect small (10--140 m) asteroids that are on a collision course with Earth. ATLAS scans the entire sky accessible from Hawaii every few days using fully robotic 0.5m f/2 Wright Schmidt telescopes located on the summit of Haleakal\=a and at Mauna Loa Observatory. Each telescope has a 5.4$\times$5.4 degree field of view with 1\farcs{}86 pixels, and during normal operations each telescope obtains four 30-second exposures of 200--250 target fields per night. This allows the two telescopes together to cover roughly half of the accessible sky per night, with the four observations of a given field typically obtained within less than an hour.  The ATLAS telescopes use two broad filters: the `cyan' filter ($c$) covering 420--650 nm and the `orange' filter ($o$) covering 560--820 nm \citep{magnier16,tonry18}.

Every image from the ATLAS telescopes is processed by a fully automated pipeline that performs flat fielding, astrometric calibration, and photometric calibration. A low-noise reference image constructed by stacking multiple images of the appropriate field taken under excellent conditions is then subtracted from each new image, allowing the detection and discovery of asteroids and other transient sources.

We performed forced photometry on the subtracted ATLAS images of ASASSN-18bt as described in \citep{tonry18}. 
We then took a weighted average of the intra-night photometric observations to get a single flux measurement for each night of observation. The ATLAS photometry and 3-sigma limits are presented in Table \ref{tab:phot} and are shown in Figure~\ref{fig:phot}.

\section{Characterizing the Early Light Curve}
\label{sec:K2_fit}

The high cadence and photometric precision of {\em Kepler}  gives us an extremely well-sampled early light curve, allowing us to fit and model the physical parameters of the supernova with a high degree of accuracy. To get a more realistic estimate for the point-to-point errors, we measure the mean and standard deviation in the {\em K2}  light curve from the beginning of Campaign 16 until 5 days before there is any signature of ASASSN-18bt in the light curve. We take that to be the point-to-point error for the entire {\em K2}  light curve. This method cannot account for any systematic errors that are coherent in time.  

\begin{figure*}
\begin{center}
\includegraphics[width=0.47\textwidth]{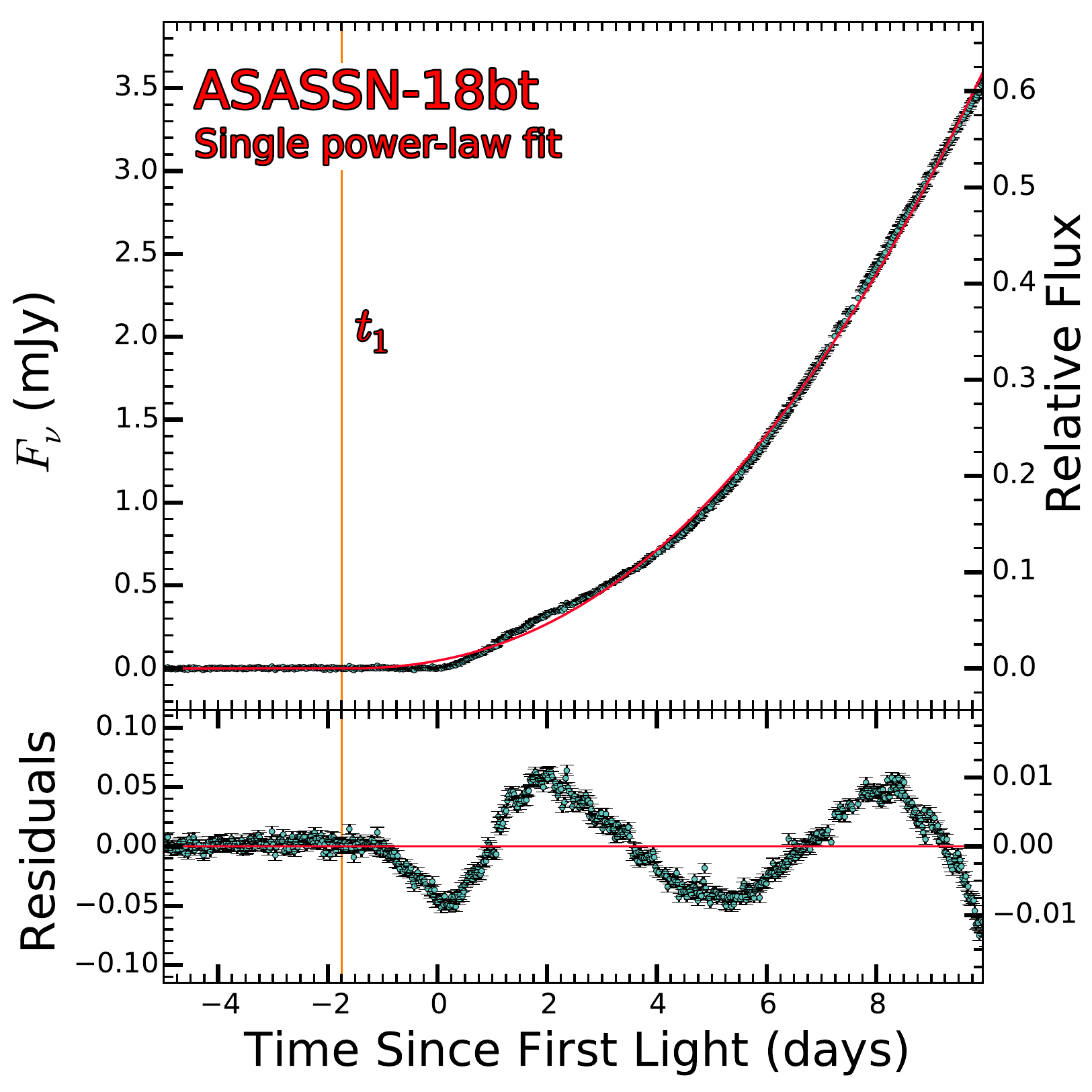}
\includegraphics[width=0.47\textwidth]{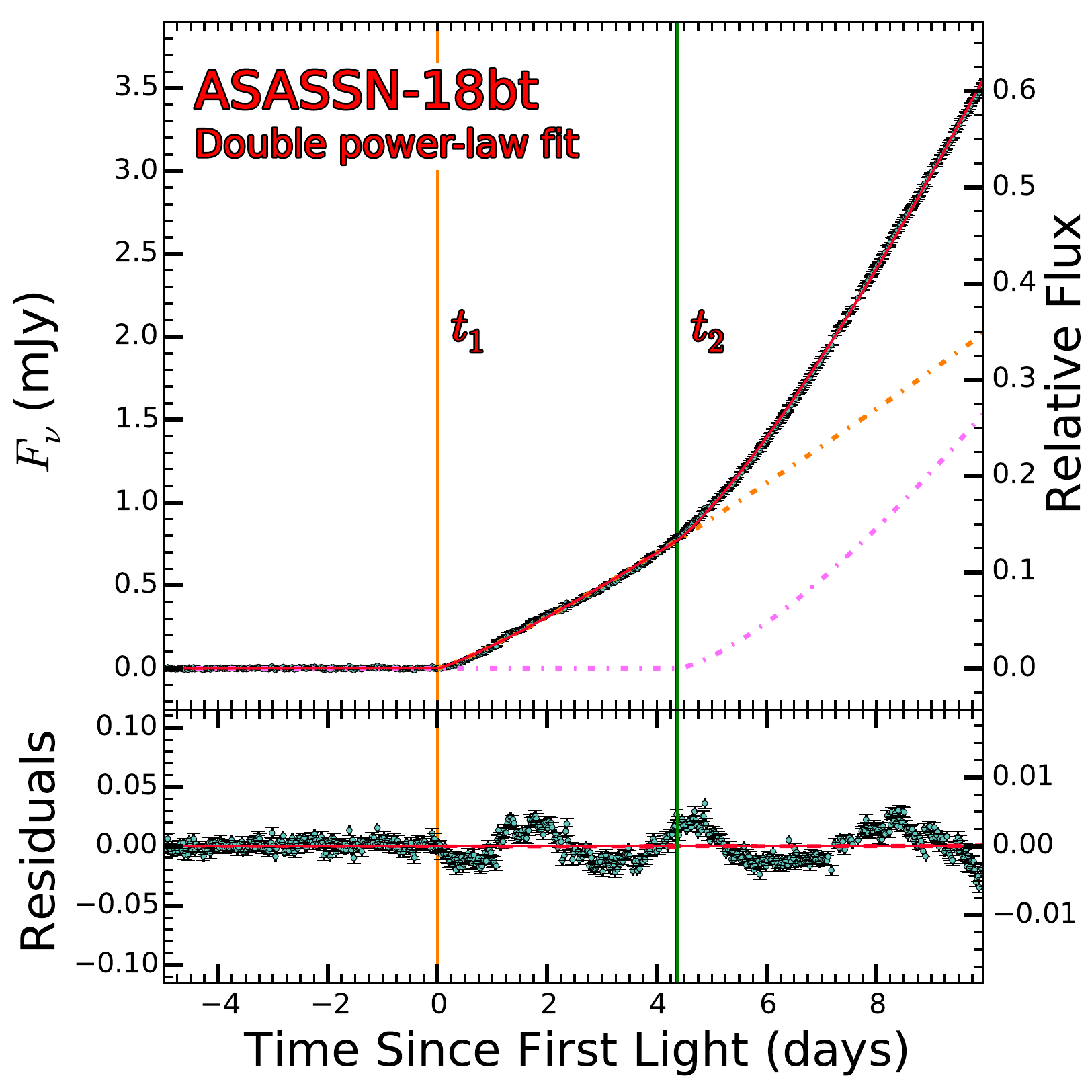}
\end{center}
\caption{The {\em K2}  early-time light curve of ASASSN-18bt and the corresponding best-fit single power-law (\emph{left panel}) and double-power-law models (\emph{right panel}). \emph{Top:} {\em K2}  flux relative to maximum brightness. The red line shows the best fit of Equation~\ref{eq:double} to the {\em K2} light curve.  The red-dashed lines indicate the 1-sigma error on the fit but are mostly underneath the solid red line.  The orange and pink dot-dashed lines show the two components of the fit. \emph{Bottom:} Residuals from the models.  The vertical orange and green lines indicate $t_{\rm{1}}$ and $t_{\rm{2}}$, respectively.}
\label{fig:K2lc}
\end{figure*}

As seen in the left panel of Figure~\ref{fig:K2lc}, it is obvious that a single power law with an arbitrary power-law index ($\alpha$) cannot adequately describe the light curve.  This also rules out an expanding fireball model where flux is proportional to a specific, $(t-t_{\textrm{1}})^{2}$, power law \citep{arnett82}. Thus, ASASSN-18bt joins a growing sample of SNe Ia with some structure in their early light curves that cannot be described by a single power-law model. It is interesting to ask what causes this structure, but first it must be characterized.  

To do so, we fit the {\em K2}  light curve with a double power law of the form
\begin{eqnarray}
\label{eq:double}
    f = z \;\; \textrm{when} \;\;  t < t_{\rm{1}}, \nonumber \\*
    f = z + h_1 ( t - t_{\rm{1}} )^{\alpha_1} \;\; \textrm{when} \;\; t_{\rm{1}} \leq t < t_2, \\*
    f = z + h_1 ( t - t_{\rm{1}} )^{\alpha_1} + h_2 ( t - t_2 )^{\alpha_2} \;\; \textrm{when} \;\; t_2 \leq t, \nonumber
\end{eqnarray}
using the {\tt emcee} Markov Chain Monte Carlo package \citep{foreman13}.  Figure~\ref{fig:K2lc} shows the {\em K2}  light curve and the best-fitting double-power-law model (top panel), as well as the fit residuals (bottom panel). The double power law describes the rising {\em K2}  light curve well with just 7 free parameters.  The pattern in the residuals is likely not due to the sawtooth thruster firing described in Section~\ref{sec:K2_LC} because the residuals are mostly symmetric in time and occur over too long of a period. Thus, the residuals likely indicate that there is some behavior not completely captured by our double-power-law model.  However, the reasonable fit and two different time scales in Equation~\ref{eq:double} imply that there may be two different physical processes contributing to the light curve. We will explore potential physical models in the next few sections. 

To estimate the peak flux and the time of maximum in the {\em Kepler}  band pass, we fit a quadratic function to the {\em K2}  light curve within $2$ days of the peak. This allows us to scale the light curve shown in Figure~\ref{fig:K2lc} to the peak and to compute the rise time in the {\em K2}  filter alone which is important when comparing to the previous SNe Ia observed by {\em Kepler} .  From the double-power-law fit we find that $t_{\textrm{1}} = \tAbt$ and with the quadratic fit to peak we find a rise time of $t_{\textrm{rise}}=t_{\textrm{peak}}-t_{\textrm{1}}=\triseKbt$~days. Throughout this work, we use this best-fit estimate of $t_{\textrm{1}}$ as the temporal origin.

\begin{figure*}
\begin{center}
\includegraphics[width=0.31\textwidth]{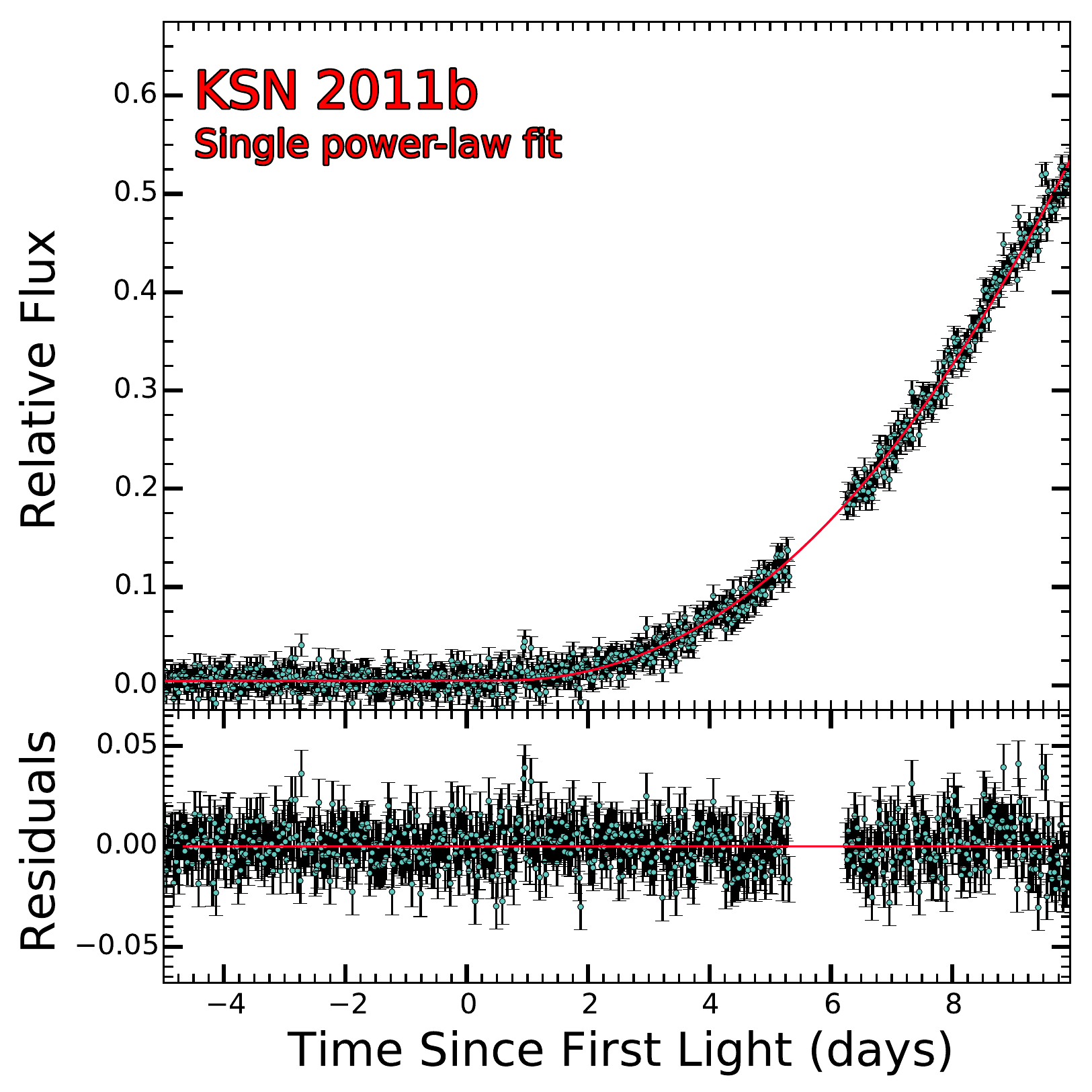}
\includegraphics[width=0.31\textwidth]{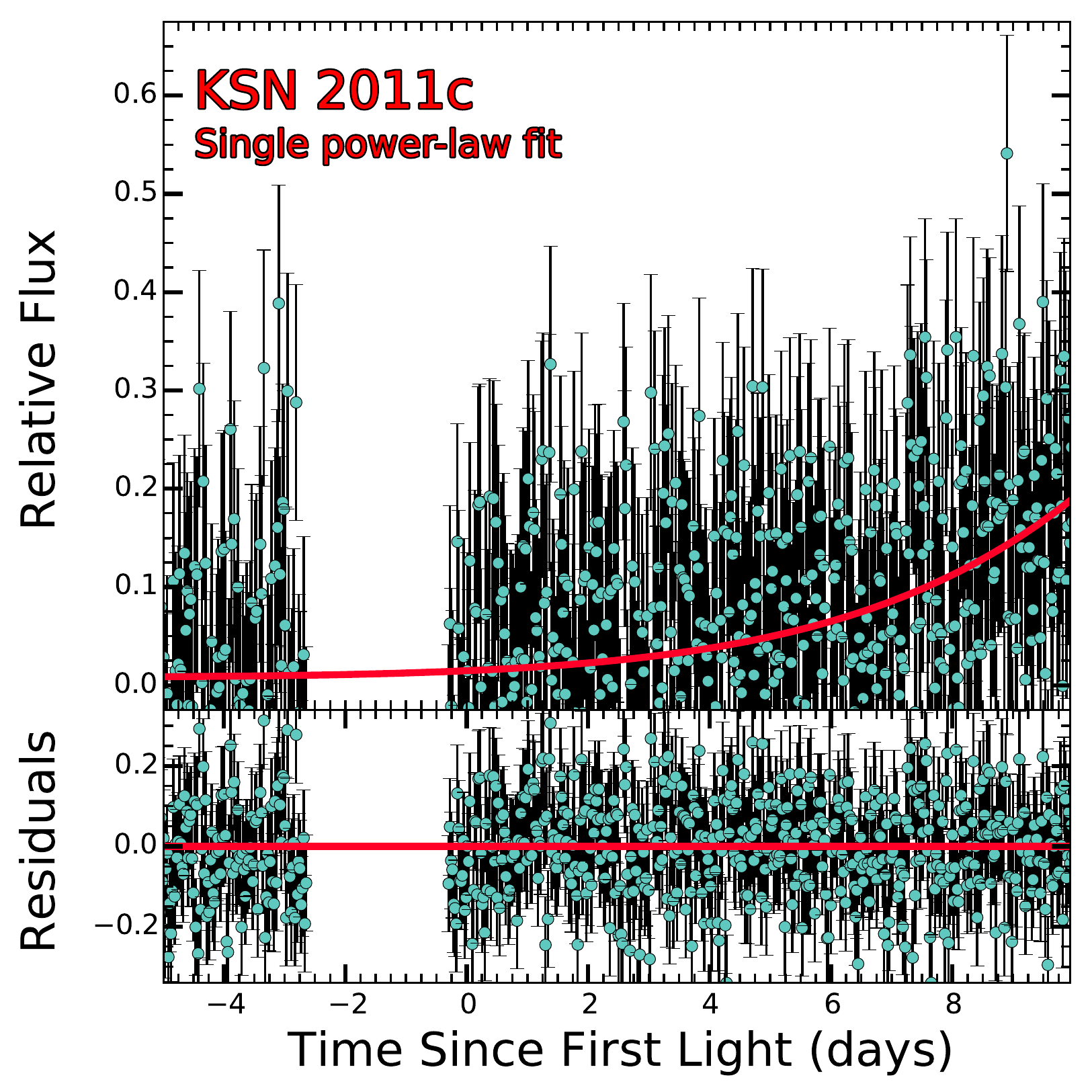}
\includegraphics[width=0.31\textwidth]{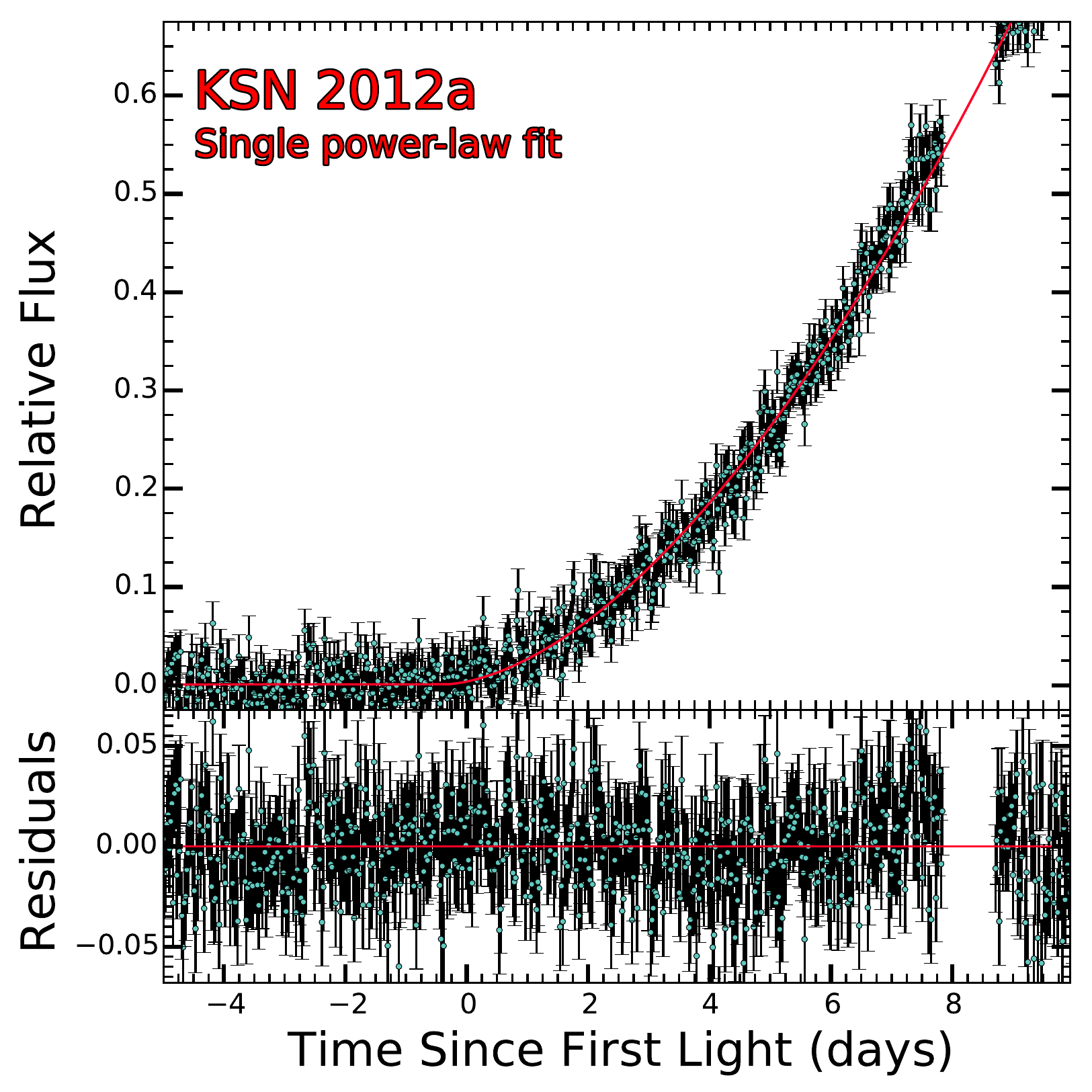}
\end{center}
\begin{center}
\includegraphics[width=0.31\textwidth]{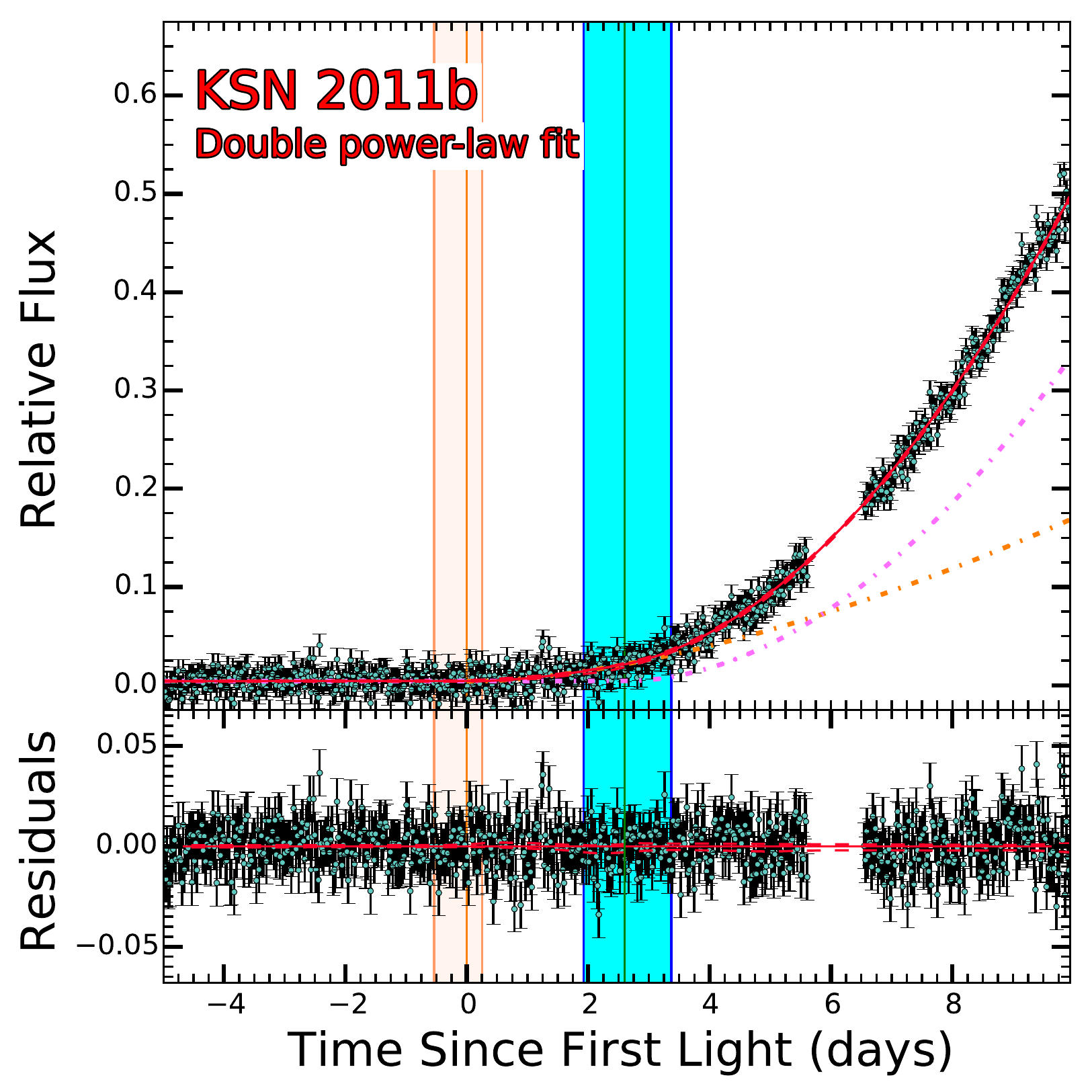}
\includegraphics[width=0.31\textwidth]{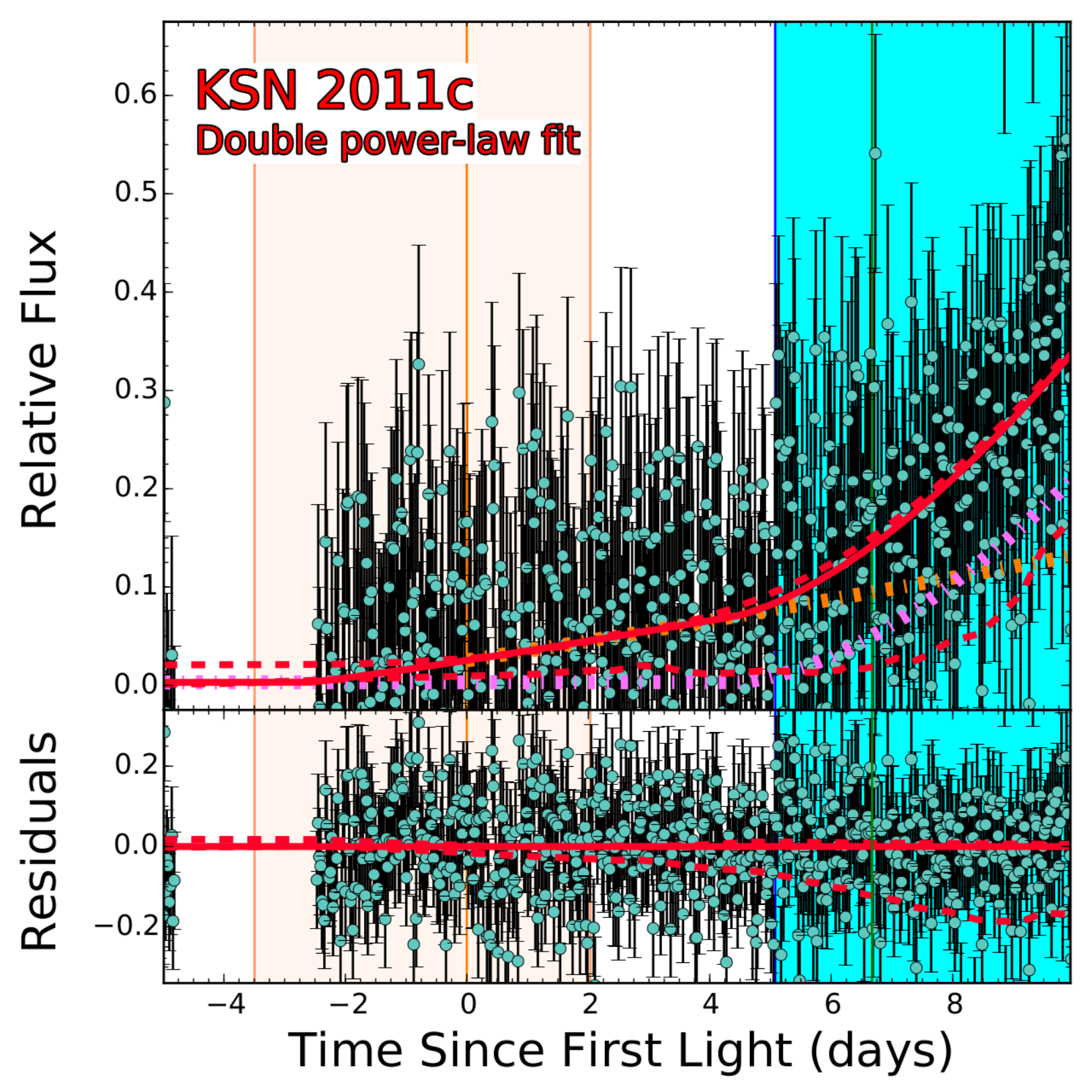}
\includegraphics[width=0.31\textwidth]{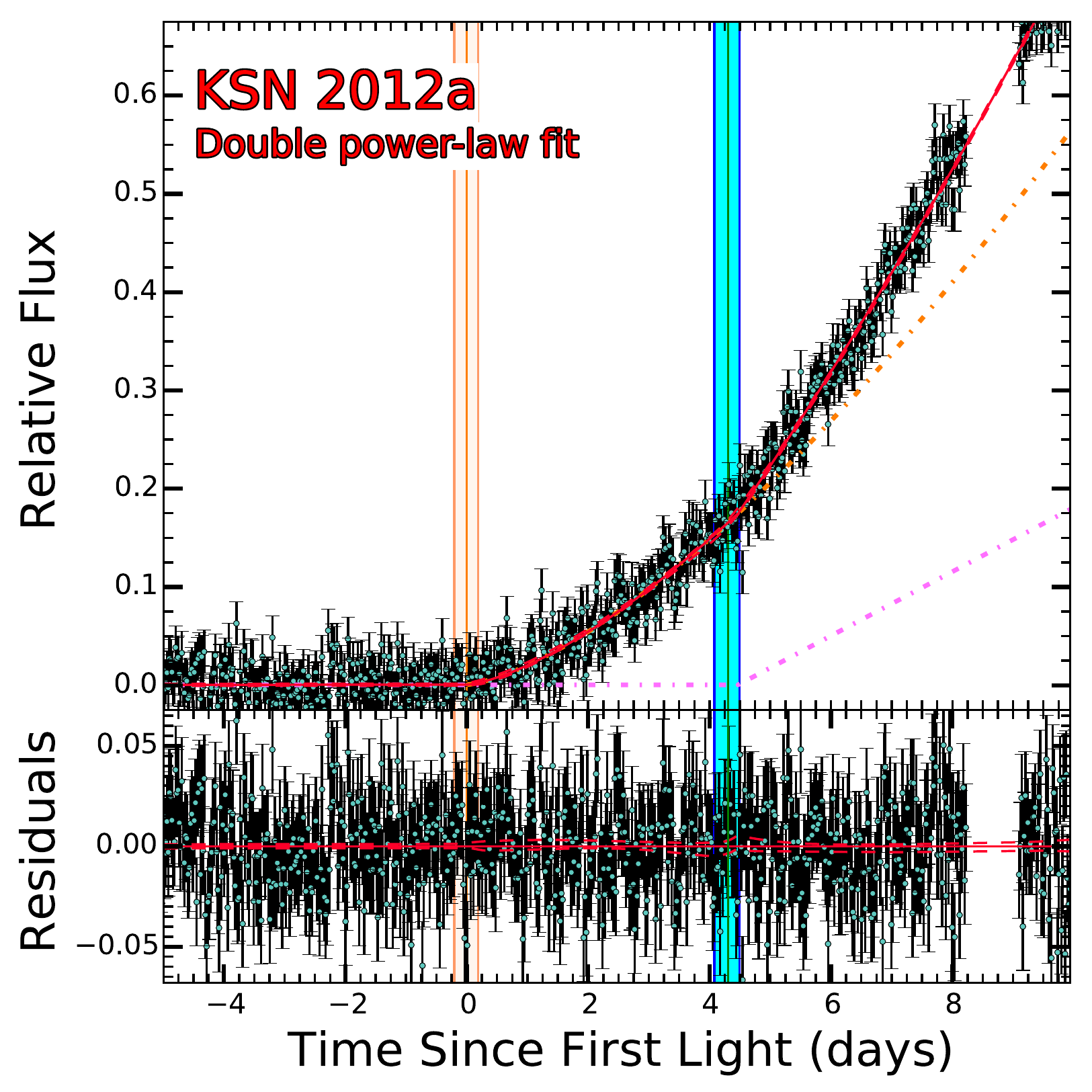}
\end{center}
\caption{{\em Kepler}  light curves and best-fit single (\emph{upper row}) and double-power-law models (\emph{bottom row}) for the other 3 SNe~Ia observed with {\em Kepler}  \citep{olling15}. Colors have the same meanings as in Figure~\ref{fig:K2lc} and the supernova names are given in the top-left corner of each panel. The light curves are plotted on the same scale as in Figure~\ref{fig:K2lc} to enable comparison between the four SNe.  While the residual panels are individually scaled for each supernova to enable a comparison of the quality of the fits. The second power law is not constrained for KSN 2011c due to the nosier data.}
\label{fig:Keplerlcs}
\end{figure*}

We also fit the previous 3 SNe Ia observed by {\em Kepler}  \citep{olling15} with the same double and single power-law models. Figure~\ref{fig:Keplerlcs} shows the these light curves and their corresponding best fits. In order to facilitate comparison among the four {\em Kepler}  SNe~Ia, Figure~\ref{fig:Keplerlcs} uses the same scale as Figure~\ref{fig:K2lc}.  The best-fit parameters from Equation~\ref{eq:double} are shown for all four SNe in Table~\ref{tab:fit}. In the Table, $t_{\textrm{rise}}$ is the time from $t_{\textrm{1}}$ to the maximum in the {\em K2}  filter ($t_{\textrm{peak}}$), while $t_{B\textrm{rise}}$ is the time from $t_{\textrm{1}}$ to the estimated time of $B$-band maximum light.  All three objects can be nearly equally well described by either a single or double-power-law fit and there is no compelling evidence that KSN 2011b, KSN 2011c, or KSN 2012a light curves require the second power-law component.  However, the light curves of all three SNe are substantially noisier, which would mask early-time behaviors.  To demonstrate this, we determine the earliest time ($t_{\textrm{det}}$) the SN light curve is 1 sigma above the average pre-explosion flux.  Of the four {\em Kepler}  SNe, only ASASSN-18bt is confidently detected within the first day of $t_{\textrm{1}}$.

\begin{deluxetable*}{clllllll}
\tablewidth{440pt}
\tabletypesize{\footnotesize}
\tablecaption{Photometric Observations}
\tablehead{
\colhead{SN} &
\colhead{$t_{\textrm{1}}$} &
\colhead{$t_2 - t_{\textrm{1}}$} &
\colhead{$\alpha_{1}$} & 
\colhead{$\alpha_{2}$} &
\colhead{$t_{\textrm{rise}}$} &
\colhead{$t_{B\textrm{rise}}$} &
\colhead{$t_{\textrm{det}} - t_{\textrm{1}}$} \\ 
\colhead{} &
\colhead{JD} &
\colhead{days}  &
\colhead{} &
\colhead{} &
\colhead{days} &
\colhead{days} &
\colhead{hours} }
\startdata
ASASSN-18bt & \tAbt   & \tBbt   & \aAbt   & \aBbt   & \triseKbt   & \triseBbt   & \tdetbt   \\
KSN 2011b   & \tAKSNb & \tBKSNb & \aAKSNb & \aBKSNb & \triseKKSNb & \triseBKSNb & \tdetKSNb \\
KSN 2011c   & \tAKSNc & \tBKSNc & \aAKSNc & \aBKSNc & \triseKKSNc & \triseBKSNc & \tdetKSNc \\
KSN 2012a   & \tAKSNa & \tBKSNa & \aAKSNa & \aBKSNa & \triseKKSNa & \triseBKSNa & \tdetKSNa 
\enddata \tablecomments{Fit parameters of the double-power-law model (Equation~\ref{eq:double}) for the 4 SNe Ia observed with {\em Kepler}  to-date. A second power law is not constrained for KSN~2011c, likely because its light curve is significantly noisier due to its greater distance.} 
\label{tab:fit} 
\end{deluxetable*} 

Next we explore some of the physical processes that could be responsible for the double-power-law structure in the early light curve of ASASSN-18bt.

\section{Early-time Light Curve and Companion Constraints}
\label{sec:companions}

If the progenitor of an SN~Ia is a WD accreting from a non-degenerate companion, then its ejecta are expected to interact with the companion after explosion, potentially producing an imprint on the early, rising light curve. The strength of this signature is thought to depend on the viewing angle with respect to the progenitor system, with the strongest effect occurring when the companion lies along the line of sight between the observer and the supernova. The effect scales proportionally with the radius of the companion, $R_\textrm{c}$, when the viewing angle is fixed. In this Section we compare the early rise of ASASSN-18bt with emission models derived for the interaction between SN~Ia ejecta with different sized companions, in order to investigate whether interaction with a companion can explain the double-power-law structure in the light curve and to place limits on $R_\textrm{c}$.  We used the analytic models from \citet{kasen10} to generate light curves for a variety of $R_\textrm{c}$ assuming the companion is aligned with our line of sight where the signature is expected to be largest.  We also assumed that the companion was Roche-lobe overflowing and that the mass of the primary and companion are 1.4 and 1.0 \msun, respectively.  This introduces a weak dependency on mass \citep{eggleton83}, but the mass dependence is unimportant compared to the unknown viewing angle.

First we simply compared the \citet{kasen10} models to our early time data assuming that the time of explosion ($t_{\textrm{exp}}$) was the same as the  $t_{\textrm{1}}$ measured from the double-power-law fit in Section~\ref{sec:K2_fit}.  While $t_{\textrm{exp}}$ and $t_{\textrm{1}}$  have occasionally been used interchangeably they need not be the same because of a possible dark phase between the explosion and when the supernova first starts to brighten \citep{hachinger13, piro14, piro16}. \citet{piro16} showed that even in extreme cases, dark phases last $< 2$ days, and more realistically last $\lesssim$ 1 day.  This effect will be discussed more in Section~\ref{sec:NiDist}.  

\begin{figure*}
\begin{center}
\includegraphics[width=0.31\textwidth]{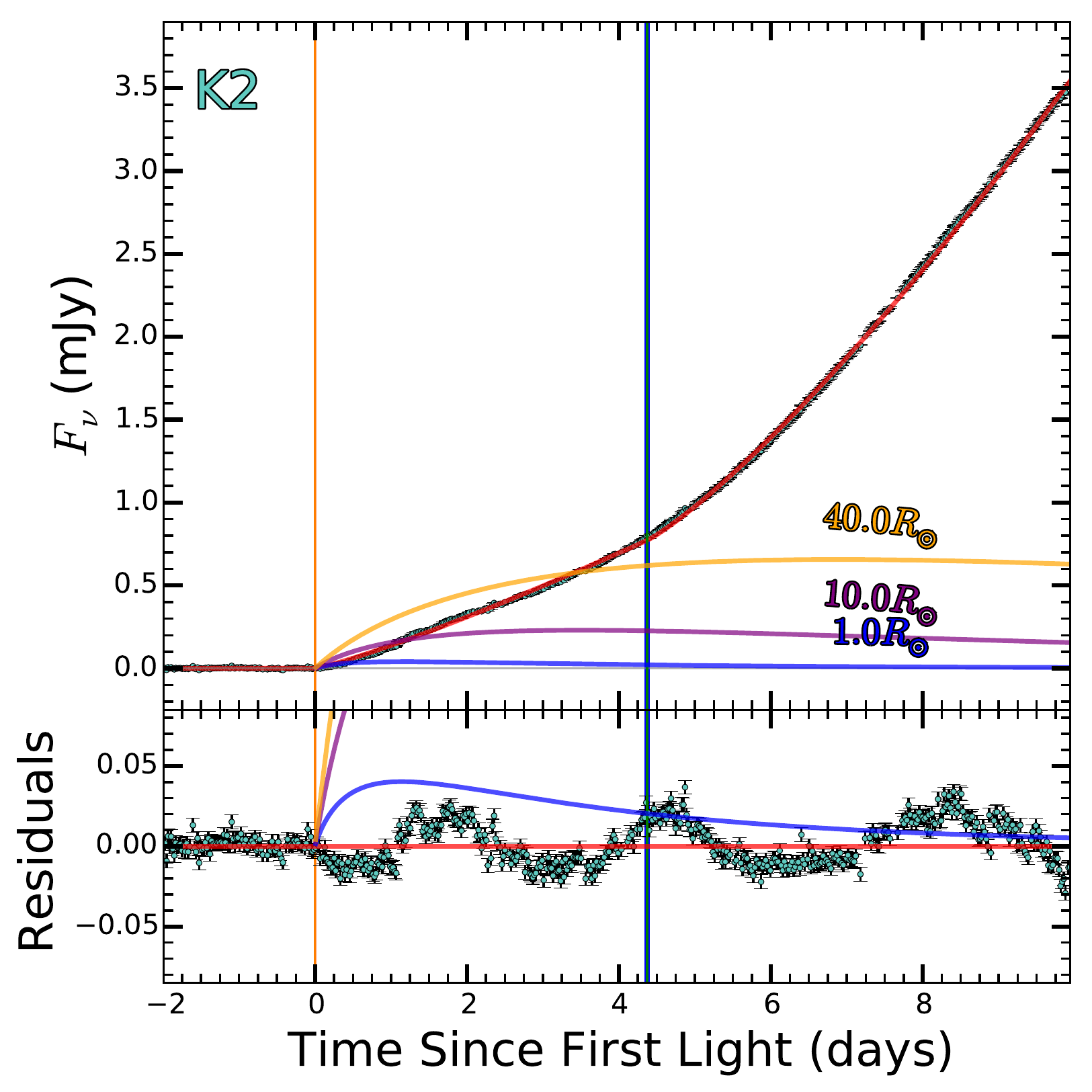}
\includegraphics[width=0.31\textwidth]{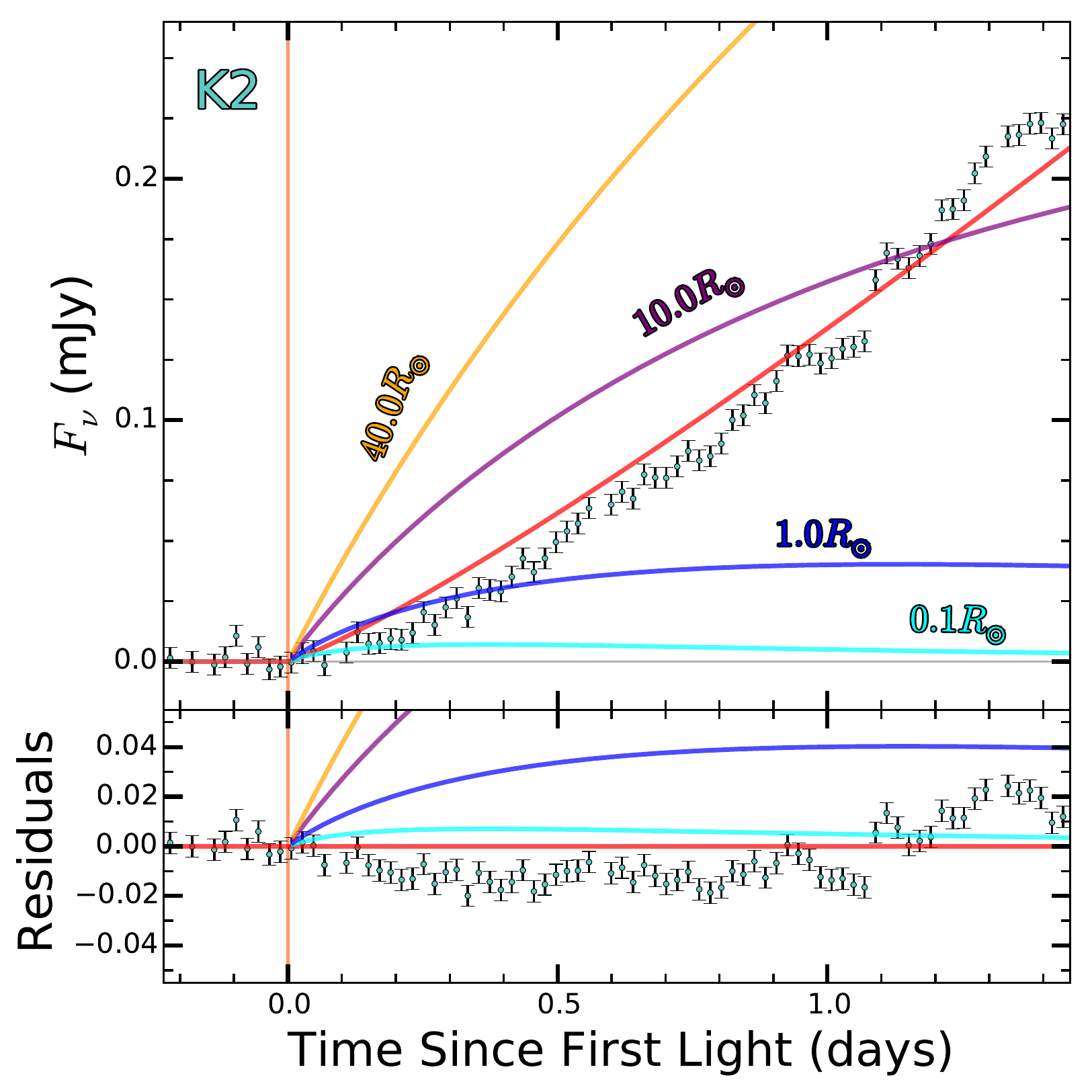}
\includegraphics[width=0.31\textwidth]{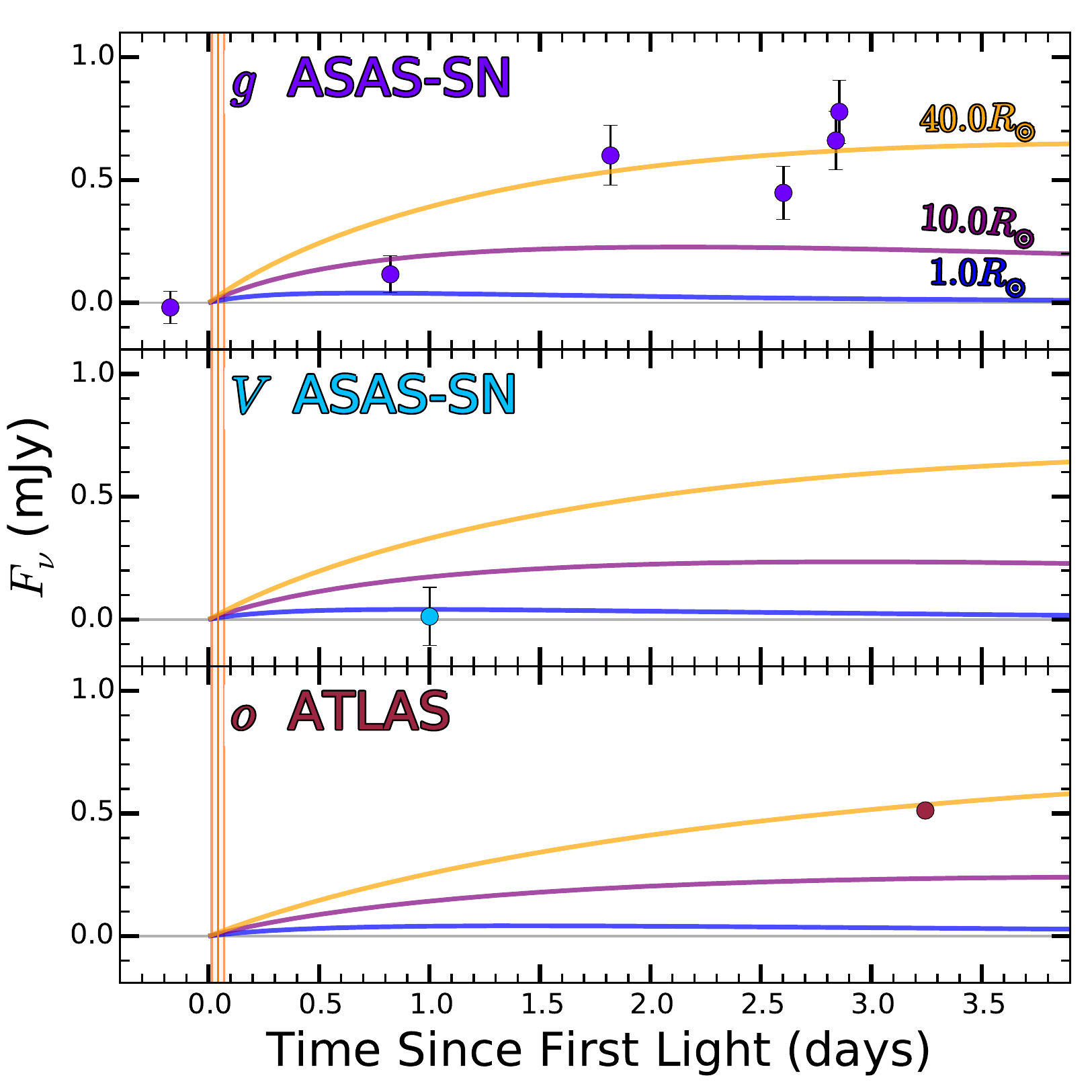}
\end{center}
\begin{center}
\includegraphics[width=0.31\textwidth]{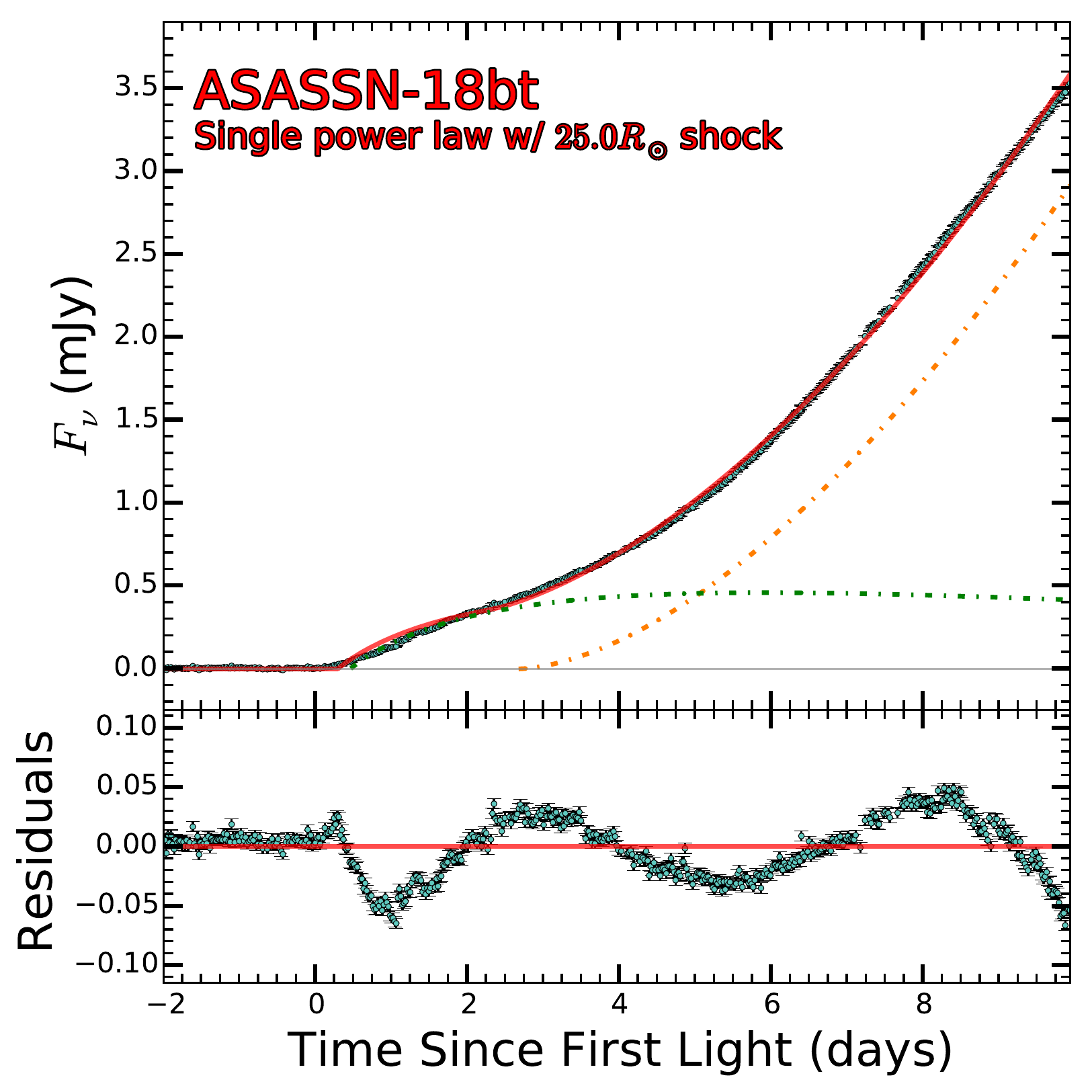}
\includegraphics[width=0.31\textwidth]{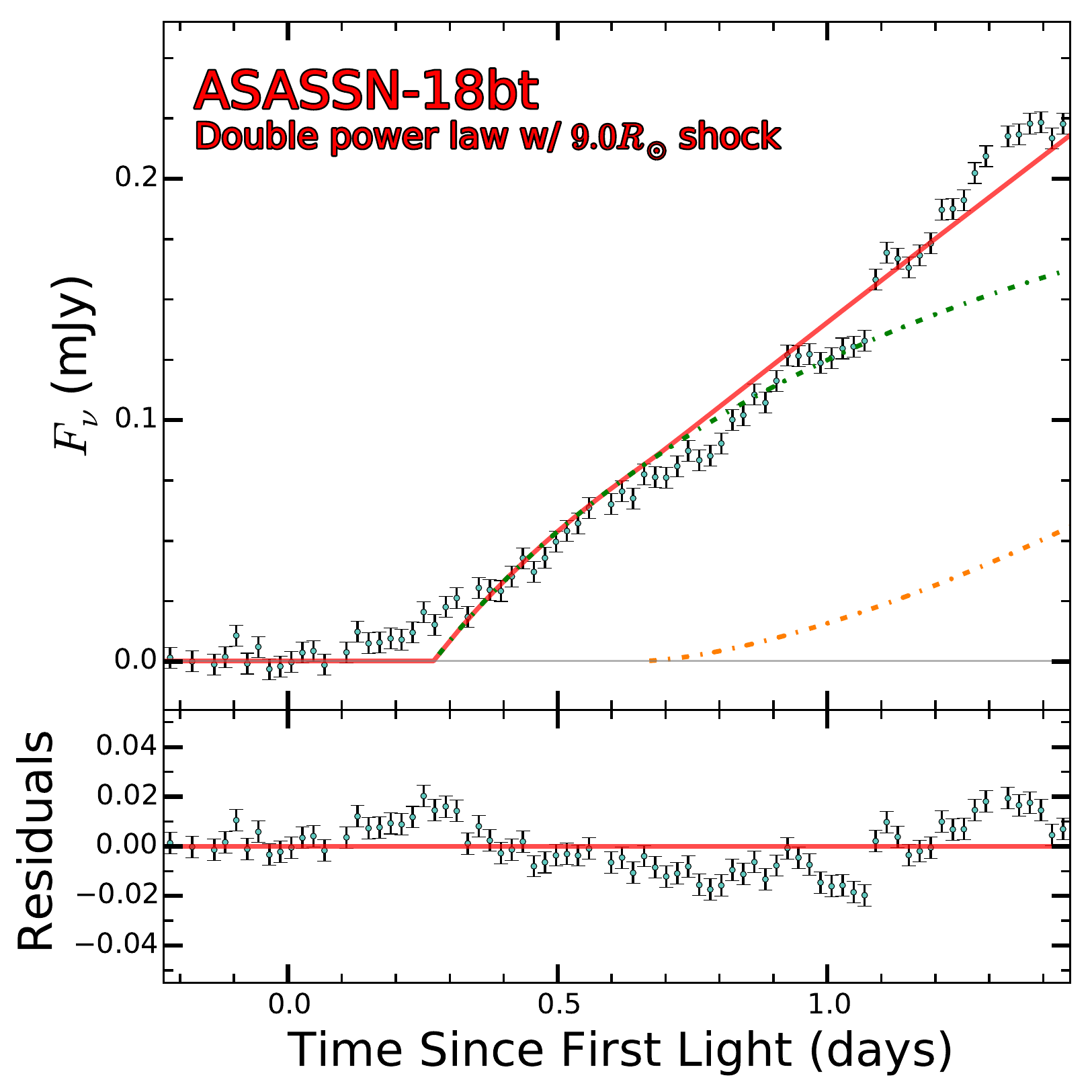}
\includegraphics[width=0.31\textwidth]{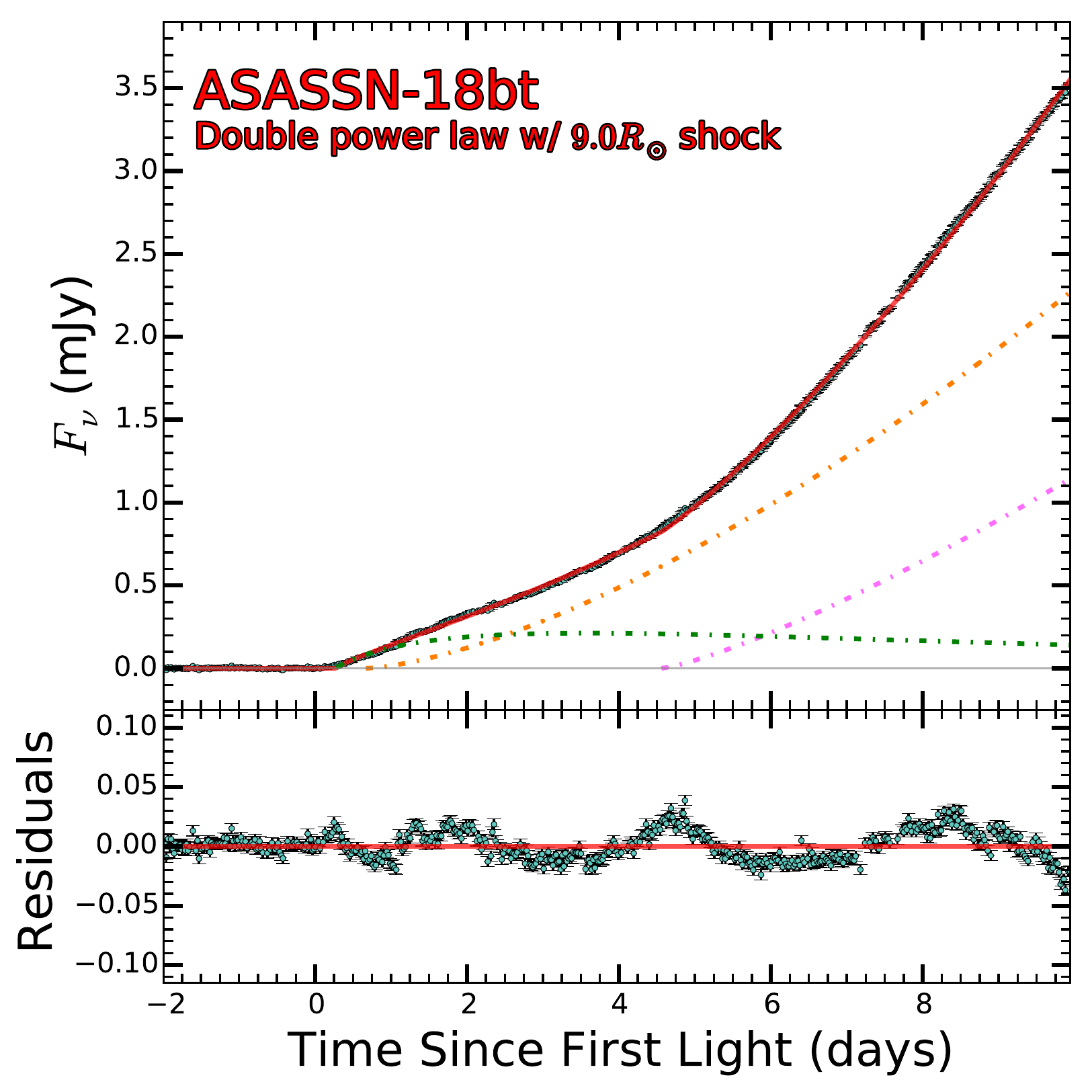}
\end{center}
\caption{(\emph{Top row:}) {\em K2} , ASAS-SN and ATLAS light curves of ASASSN-18bt compared to the \citet{kasen10} models of emission from the interaction of the supernova shock with companions of various radii assuming the companion is along our line of sight. The left and center panels shows the first 10 and 1.5 days following $t_{\textrm{1}}$. The right panel shows the ASAS-SN and ATLAS light curves. (\emph{Bottom row}) The left panel shows the {\em K2}  light curve fit with a best-fit single power-law and companion model. It can be seen that a single power-law and companion model cannot satisfactorily reproduce the observed light curve. The center and right panel show the largest radius companion allowable with a double-power-law fit. See Section~\ref{sec:companions} for details.  }
\label{fig:companion}
\end{figure*}

In the top row of Figure~\ref{fig:companion} we compare the early light curves from {\em K2} , ASAS-SN and ATLAS to the interaction models for a 0.1, 1.0, 10.0 and 40.0 \rsun{} companion. In the upper left panel it can be immediately seen that if the initial nearly-linear rise is to be explained by the interaction with a companion, it must be a large companion ($\sim 40$ \rsun) to produce a large enough signature.  However, the upper center and upper right panels show that the early {\em K2} , ASAS-SN $g$-band, and ASAS-SN $V$-band light curves are inconsistent with such a large signature from a companion and immediately rule out companions significantly larger than $\sim 10 \rsun$ for our assumed viewing angle.  

To further demonstrate that the early time light curve of ASASSN-18bt cannot be described by a single power-law rise combined with an interaction with a companion we construct a grid of companion models and simultaneously fit the companion radius and power-law component.  The best-fit model is shown in the bottom left panel of Figure~\ref{fig:companion}.  The best-fit companion radius is 25 \rsun{} but the fit has large residuals. The main issue is that the interactions produce a light curve that rises rapidly and then flattens while the observed light curve rises nearly linearly and then steepens (see Table~\ref{tab:fit}).  As discussed in the previous paragraph such a large companion is also inconsistent with the bluer ASAS-SN pre-discovery data.  Thus, if an interaction with a companion contributes significantly to the rise of ASASSN-18bt, the intrinsic rise of the SN itself must be more complicated than a single power law. 

Next we simultaneously fit for a companion radius and a double-power-law model (Equation~\ref{eq:double}).  We constrained the dark time to ($t_{\textrm{1}} - t_{\textrm{exp}}$) to be positive, assuming the progenitor cannot emit significant flux prior to explosion, and less than 2.0 days.  Additionally, we constrained $h_1$ and $h_2$ to be positive and $a_1$ and $a_2$ to be greater than 1. Finally, we constrained $t_{\textrm{1}}$ to be within 0.3 days of $t_{\textrm{det}}$ as measured in Section~\ref{sec:K2_fit}. 

We find that the first power-law component and the companion can compensate for each other and that the dark time, the power-law index, and the companion radius are degenerate because the \citet{kasen10} companion models initially rise quickly and then turn over in the {\em K2}  filter whereas any power law with $\alpha_1 > 1$ does the opposite.  Thus a nearly linear rise is possible in the first $\sim4$ days without any strong kinks or features.  {\emph This, however, requires fine tuning of the power law and companion to hide the shock signature in a smooth curve.} Although, strictly speaking, solutions can be found.

To place a statistical limit on the radius of a companion assuming the rise can be well described by a double-power-law model, we first found the best fit for companion models from 0.01 to 50.0 \rsun{}.  We found nearly identically good fits for radii from 0.01-8 \rsun{} before the fits begin to deteriorate.  To place a statistical upper limit we focus between $-$0.5 and 2 days where a companion might contribute significantly to the light curve. We then found where the $\chisq{}$ probability distribution was $<0.32$ and $<0.05$ during that time period.  We find that the largest radii companions that have acceptable fits under these criteria are 8.0 \rsun{} and 11.5 \rsun{}, respectively. For reference, we plot the smallest companion radius ruled out at 1 sigma in the bottom center and bottom right panels of Figure~\ref{fig:companion}. It can be seen that to fit a 9 \rsun{} companion, $t_{\textrm{exp}}$ is being pushed to be later than in the fits using only a double power law and that the model misses the earliest rise of the light curve.  This weak constraint on the progenitor system demonstrates that a physically motivated model for the rising SN light curve is required before we can confidently use early time light curves of SNe Ia to constrain their progenitor systems.

\newpage     

\section{Comparison to $^{56}$Ni Mixing Models}
\label{sec:NiDist}

Very early-time emission from SNe~Ia can probe the location of  $^{56}$Ni in the ejecta \citep[e.g.,][]{piro13} and thus can be used as a diagnostic of the explosion physics. In \citet{piro16} the authors used the open-source SuperNova Explosion Code \citep[SNEC;][]{morozova15} to investigate how the distribution of $^{56}$Ni can affect the earliest phases of SN~Ia light curves. Models with $^{56}$Ni significantly mixed into the ejecta result in a quicker rise than those with  $^{56}$Ni more centrally concentrated. \citet{contreras18} matched the early light curve of SN~2012fr with model light curves predicted for different levels of $^{56}$Ni mixing. They found that the early steepening seen in the light curve of SN~2012fr could be accounted for by a model with a $^{56}$Ni mass fraction of 0.05 at approximately 0.05~\msun{} below the surface of the WD. 

We used the same $^{56}$Ni mixing models as \citet{contreras18}. However, even after appropriately re-scaling the models for Milky Way reddening, host galaxy reddening, and differences in distance, we still found that the 
\citet{contreras18} models under predicted the observed {\em K2}  light curve.  We assume that this difference is due to the modest difference in passbands between the LSQ $gr$-band used to construct the models and the {\em K2}  band pass along with differences in the total $^{56}$Ni production between the two SNe.  We found that scaling the models by 130\% brought them into reasonable agreement with the $K2$ data. 

\begin{figure*}
\begin{center}
\includegraphics[width=0.45\textwidth]{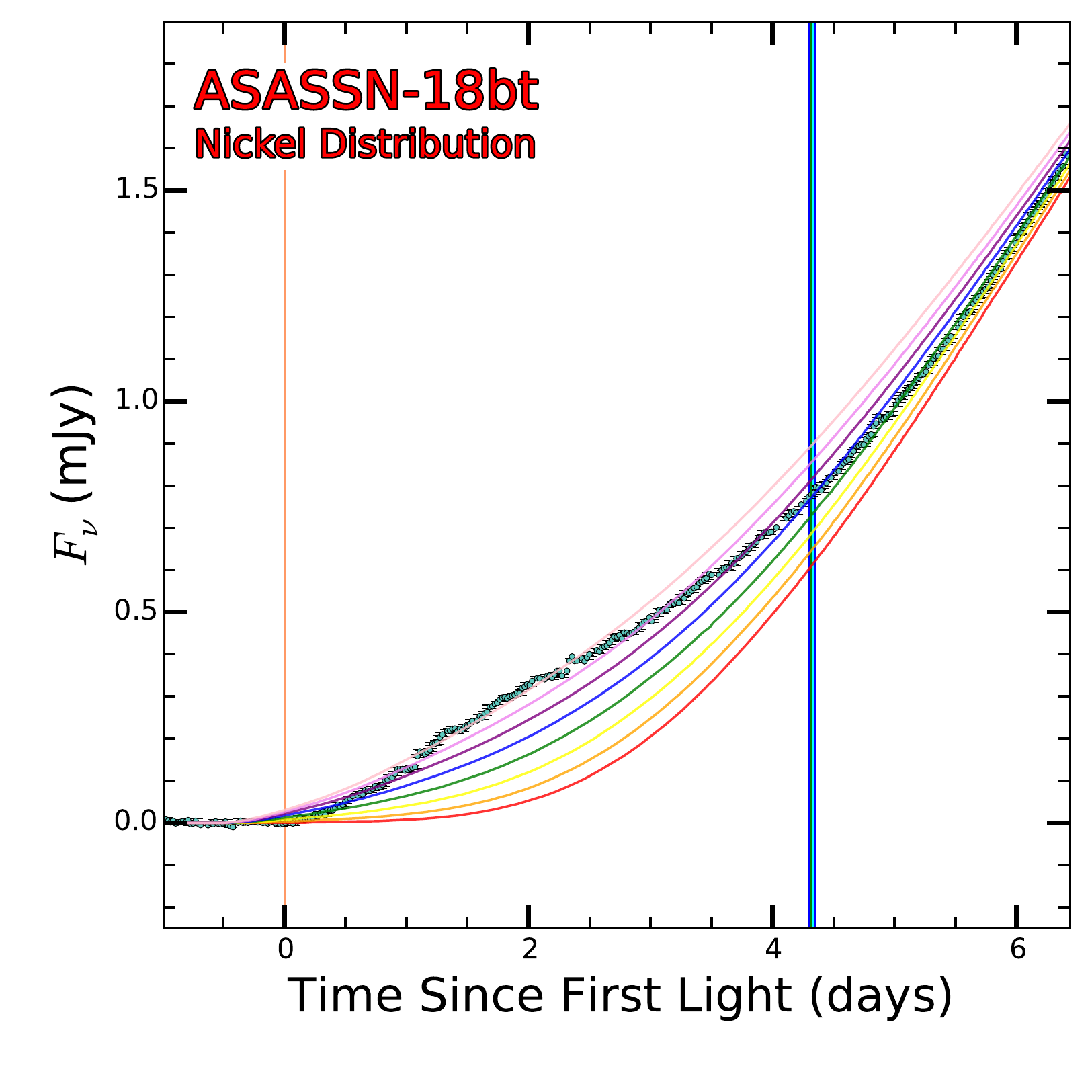}
\includegraphics[width=0.45\textwidth]{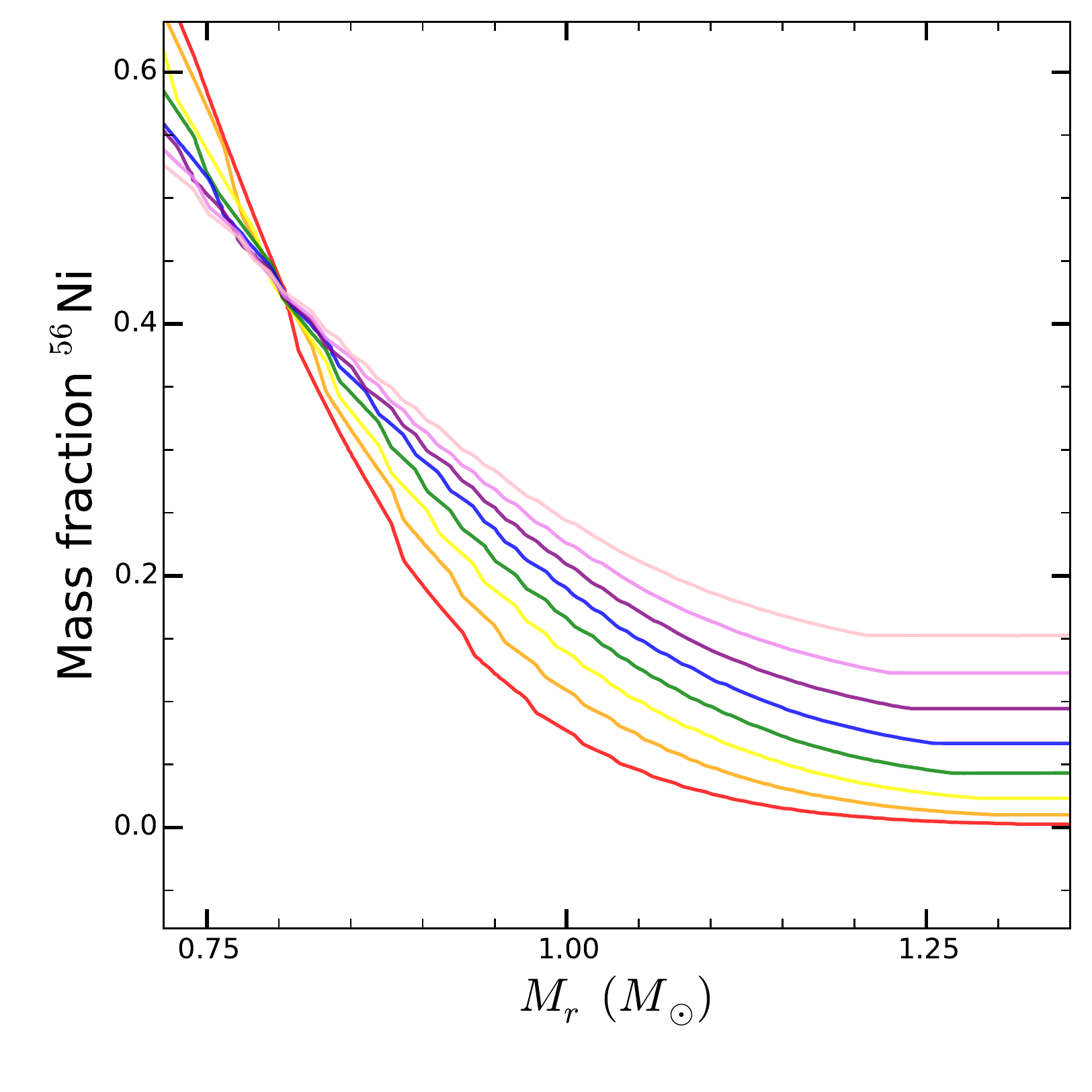}
\end{center}

\caption{\emph{Left Panel:} The scaled {\em K2}  early-time light curve of ASASSN-18bt and model light curves from \citet{contreras18} with variable $^{56}$Ni mixing. Model colors correspond to the $^{56}$Ni distributions shown in the right panel which is reproduced from \citet{contreras18}.}
\label{fig:NickleLC}
\end{figure*}

In Figure~\ref{fig:NickleLC} (left panel) we show the scaled ($\sim 60 \%$) $^{56}$Ni mixing models from \citet{contreras18}, using the same colors and scales, along with the {\em K2}  light curve of ASASSN-18bt. The right panel shows the corresponding $^{56}$Ni distributions for each model. The very early light curve is most consistent with a model where the $^{56}$Ni is significantly mixed, with a $^{56}$Ni mass fraction of $0.15-0.2$ at approximately 0.05~\msun{} below the surface of the WD. However, $\sim3$ days after first light, the light curve becomes more consistent with the moderately mixed $^{56}$Ni curves, similar to SN~2012fr. This might imply that the $^{56}$Ni distribution in the ejecta is not smoothly varying or monotonically decreasing with radius in ASASSN-18bt.

\begin{figure*}
\begin{center}
\includegraphics[width=0.45\textwidth]{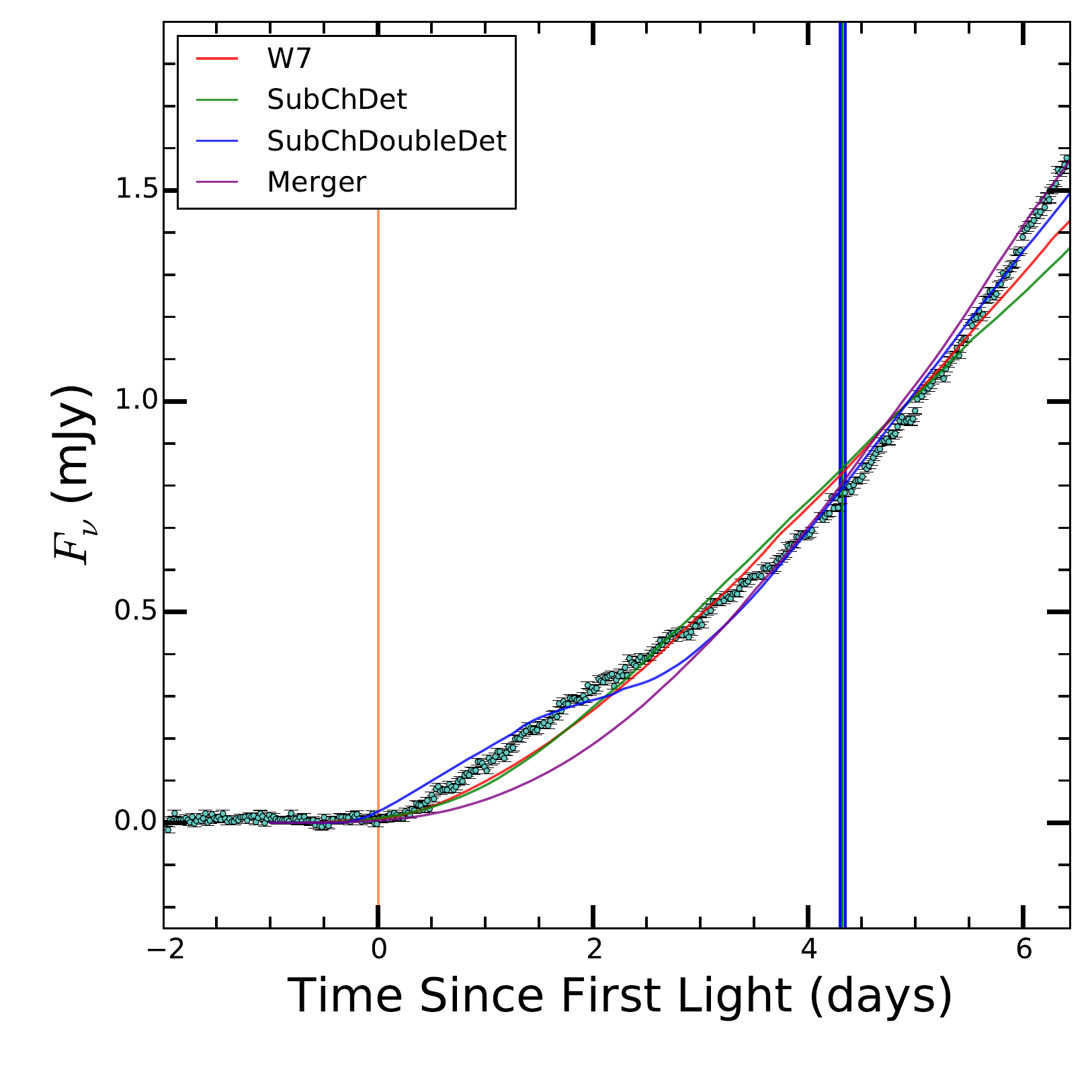}
\includegraphics[width=0.45\textwidth]{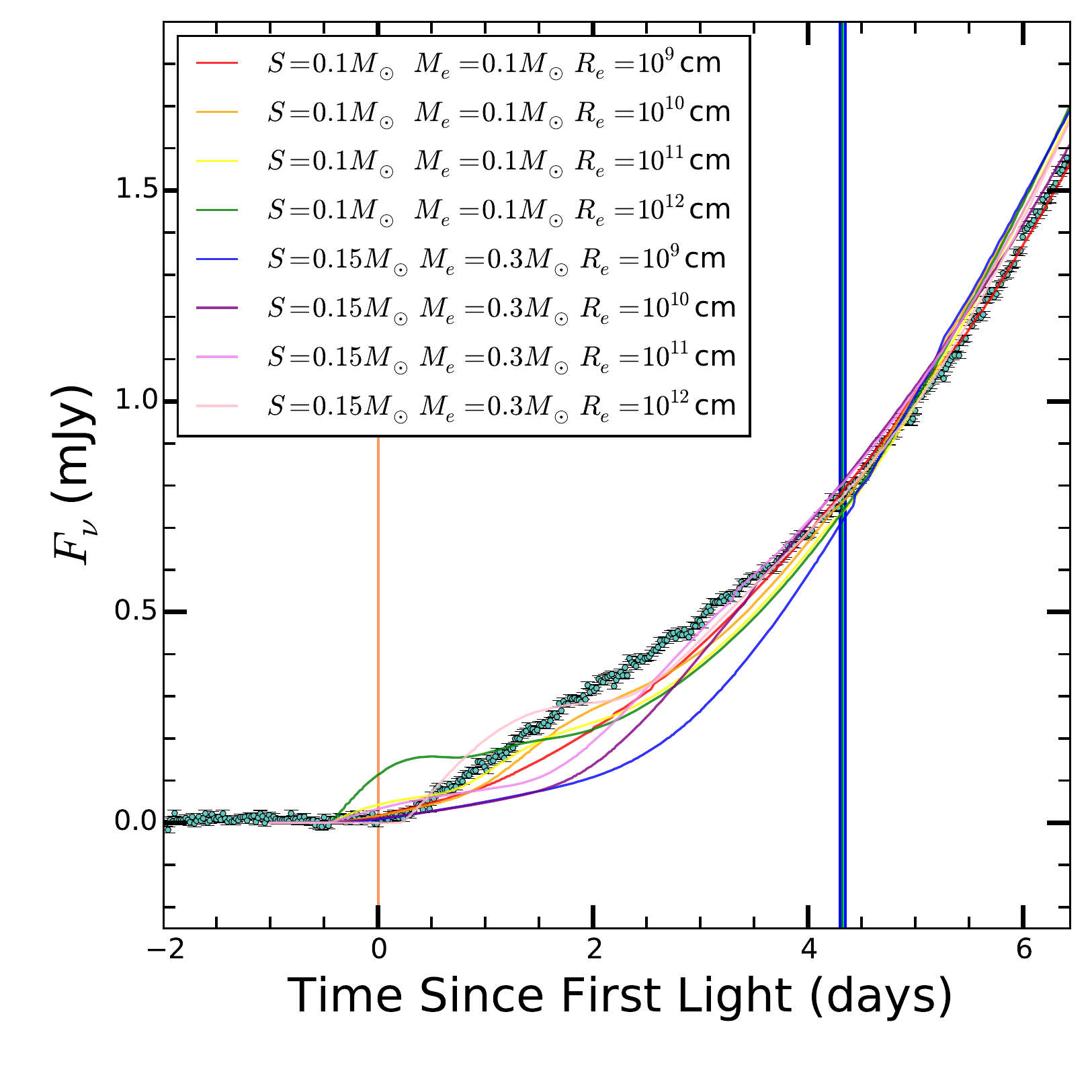}
\end{center}

\caption{The scaled {\em K2}  early-time light curve of ASASSN-18bt and model light curves.  \emph{Left Panel:} Synthetic light curves for a number of explosion models from \citet{noebauer17}.    \emph{Right Panel:} Model light curves from \citet{piro16} varying the distribution of circumstellar material and $^{56}$Ni mixing.} 
\label{fig:NickleMatLC}
\end{figure*}

Finally, in the left panel of Figure~\ref{fig:NickleMatLC} we compare the {\em K2} light curve to synthetic light curves from \citet{noebauer17}, who used the radiation hydrodynamical code Stella to compute light curves for variety of explosion models.  We compare ASASSN-18bt to the scaled predicted $V$-band light curves for 4 explosion models:\\
{\bf 1)} The parametrized 1D ejecta structure of the W7 model of \citet{nomoto84}.\\
{\bf 2)} The centrally ignited detonation of a sub-Chandrasekhar mass CO WD (SubChDet; \citealp{sim10}).\\
{\bf 3)} A "Double-detonation" model where an initial detonation in an accreted He surface layer triggers carbon detonation in the core of the sub-Chandrasekhar mass WD (SubChDoubleDet; \citealp{fink10, kromer10}). \\
{\bf 4)} The "Violent merger" of two sub-Chandrasekhar mass CO WDs, which triggers the more massive to detonate (Merger; \citealp{pakmor12}). \\
As seen in Figure~\ref{fig:NickleMatLC}, only the double-detonation model can qualitatively match the rise for the first few days. In this model, He burning leaves radioactive isotopes near the surface of the ejecta, similar to the $^{56}$Ni mixing models. Lastly, collision models (e.g., \citealp{dong15,dong18}) may also produce similar features, but the early-time light curves from this model have not, to the authors' knowledge, been investigated thoroughly.

\section{Interaction with Nearby circumstellar Material}
\label{sec:circumstellar}

The presence of a dense CSM can also affect the early time rising light curve.  As previously discussed, some SNe Ia models have a nearby non-degenerate companions but more general distributions of material are possible. Most progenitor scenarios require mass transfer, which is not a completely efficient process. \citet{piro16} investigated the possible impact of this material on the early-time light curves of SNe~Ia.  Motivated by the post merger studies of \citet{pakmor12}, \citet{shen12}, and \citet{schwab12}, \citet{piro16} argue that nearby circumstellar material is likely distributed as $\rho \propto r^{-3}$ and model the resulting light curves as a function of the total circumstellar mass ($M_e$) and its outer radius ($R_e$).  They also explore different $^{56}$Ni distributions implemented as a boxcar average with width $S$ in mass.

In the right panel of Figure~\ref{fig:NickleMatLC} we compare the  \citet{piro16} models to the {\em K2}  light curve of ASASSN-18bt. \citet{piro16} presented model $V$-band light curves whereas the {\em K2}  filter is significantly broader. We fit each model to ASASSN-18bt varying $t_{\textrm{exp}}$ and the flux scaling.  While filter differences may lead to some systematic uncertainties, we can qualitatively see that none of these models describe the data well.  All these models have trouble producing a nearly linear light curve for the first 4 days and under predict the flux around 2 days after maximum light.

\section{X-ray Limits on Progenitor Mass-loss}
\label{sec:xray}

In this section we model \swift{} X-ray observations to constrain the circumstellar material at much larger distances and lower densities.  The X-ray emission depends on both the properties of the SN, such as ejecta mass and shock velocity, and on the density of the CSM, which is sculpted by the pre-SN evolution of the progenitor system. As a result, X-ray emission offers a means to probe the nature of the progenitor system that is independent and complementary to the early light curve evolution.  The environments around SN~Ia progenitors are expected to be low-density ($\dot{M}\lesssim 10^{-9}-10^{-4}$ M$_{\odot}$~yr$^{-1}$; \citealp{chomiuk16}). Under these circumstances, Inverse Compton (IC) emission will dominate the X-ray emission at early times ($t\lesssim40$ days), when the bolometric luminosity is high \citep{chevalier06,margutti12}. 

ASASSN-18bt was observed with the \textit{Neil Gehrels Swift Gamma-ray Burst Mission} \citep[\swift;][]{gehrels04} X-ray Telescope (XRT; \citealt[][]{hill04,burrows05}) beginning on 2018-02-05 09:36:00 UTC (MJD$=58154.4$), $\sim$10 days post-explosion. In total, 10 epochs of observations were obtained over 40 days, covering the time period in which the supernova reached maximum light. All observations were reprocessed from level one XRT data using the \swift{} \textsc{xrtpipeline} version 0.13.2 script, following the standard filter and screening criteria suggested in the \swift{} XRT data reduction guide\footnote{\url{http://swift.gsfc.nasa.gov/analysis/xrt\_swguide\_v1\_2.pdf}} and the most up to date calibration files.

We inspected the individual observations and found no X-ray emission associated with the position of ASASSN-18bt. In order to place the strongest possible constraint on the presence of X-ray emission from this source we combined the individual \swift{} observations for a total exposure time of 12.6\,ks.  We again find no evidence for X-ray emission. Due to the presence of a bright X-ray point source located at $(\alpha,\delta)=(09^{h}06^{m}41.6^{s}, +19^{\circ}20'53'')$, $\sim50\arcsec$ away from the position of ASASSN-18bt, we used a source region centered on the position of ASASSN-18bt with a radius of 10$\arcsec$ combined with a standard aperture correction. 
We derive a $3\sigma$ count-rate upper limit of $2.9\times10^{-4}$ counts sec$^{-1}$ in the 0.3--10.0 keV energy band. Assuming an absorbed power law with a photon index of $\Gamma=2$, a Galactic H\textsc{i} column density of $3.42\times10^{20}$ cm$^{-2}$ derived from \citet{kalberla05}, we derive an un-absorbed flux limit of  $1.1\times10^{-14} \rm{erg\,s^{-1}cm^{-2}}$ or a luminosity of  $L_{X}(\rm 0.3-10\,keV)=3.2\times 10^{39}\,\rm{erg\,s^{-1}}$.

To constrain the density of the CSM surrounding ASASSN-18bt and thus the progenitor system mass-loss rate, we follow the same procedure as described in \citet{shappee18} for SN~2012cg. We utilize the generalized formalism developed by \citet{margutti12} for IC X-ray emission from supernovae with compact progenitors. In this formalism, the IC luminosity is directly proportional to the bolometric luminosity of the supernovae.  We adopt the bolometric light curve for ASASSN-18bt calculated in Li et al. (2018).  The deepest limits to the density of the CSM surrounding ASASSN-18bt come from the observations at $\sim$11$-$14 days post-explosion, when the bolometric luminosity was near its peak.  For a constant density CSM ($\rho_{\textrm{CSM}}=\textrm{const}$), we derive $\rho_{\textrm{CSM}}<\rm{4.5 \times 10^{5}\,\rm{cm^{-3}}}$ at a radius of 4 $\times$ 10$^{15}$ cm from the progenitor star. For a wind-like environment, the density of the CSM is $\rho_{\textrm{CSM}}=\dot M/(4\pi r^2\,v_{\textrm{w}})$, where $\dot M$ is the (constant) mass loss rate and $v_\textrm{w}$ is the wind velocity.  Following  \citet{margutti12} we find our observed X-ray flux limit implies to a mass-loss limit of $\dot M<\rm{8 \times 10^{-6}}\,\rm{M_{\sun}yr^{-1}}$ for $v_\textrm{w}=100\,\rm{km\,s^{-1}}$, at a radius of 4.5 $\times$ 10$^{15}$ cm from the progenitor star.

\begin{figure}
\includegraphics[trim={0 0 0 10cm},clip,width=0.48\textwidth]{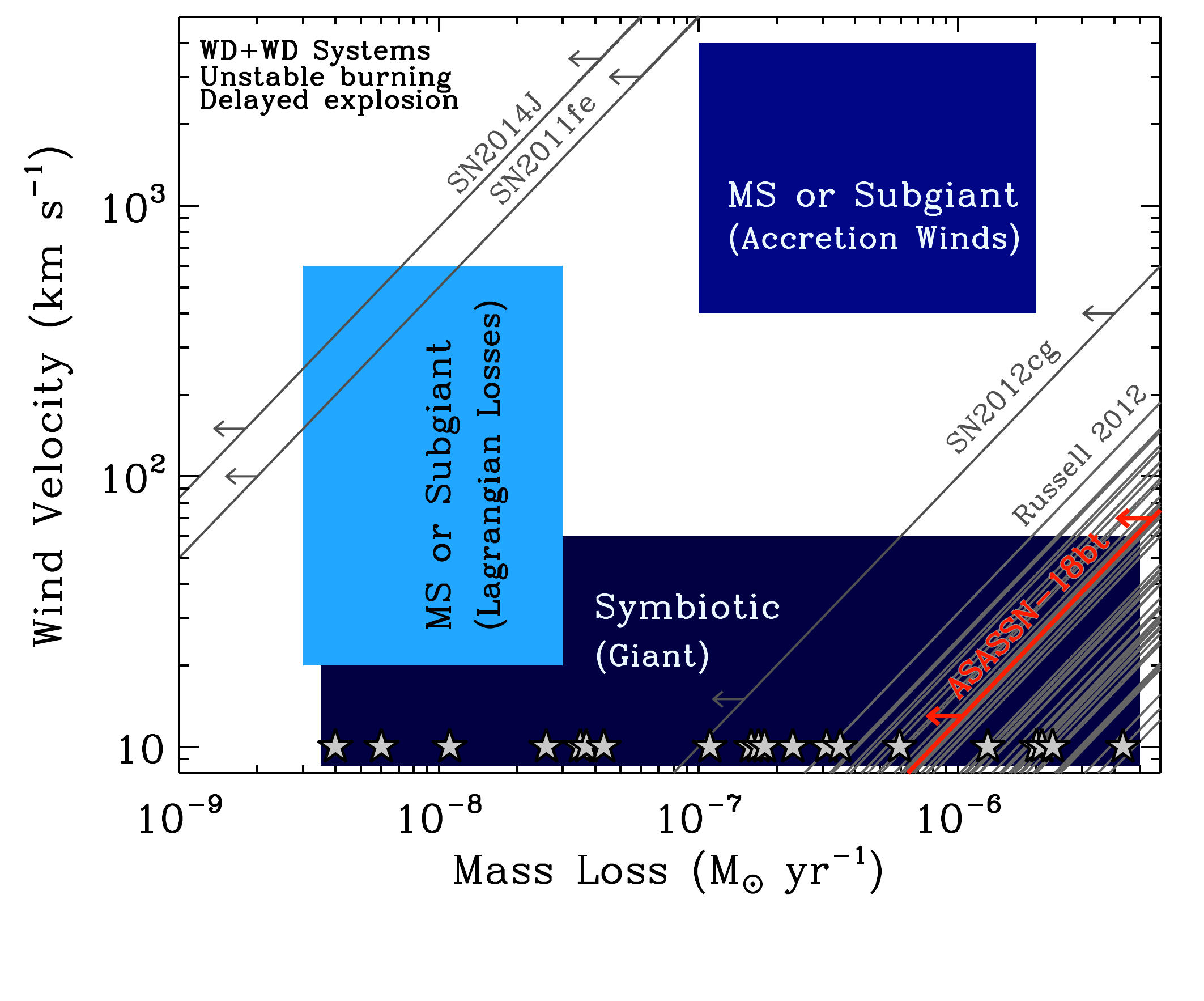}
\caption{Mass loss rate versus wind velocity. Regions occupied by a variety of proposed \snia\ progenitor systems are indicated. Diagonal lines represent limits on the progenitor mass loss rates as a function of wind velocity for observed \sneia, obtained via X-ray observations \citep{margutti12,russell12,margutti14,shappee14}. For each SN, combinations of mass loss rate and wind velocity below the line are excluded.  The limit for ASASSN-18bt, derived in Section~\ref{sec:xray}, is plotted in red. While a fraction of observed symbiotic systems are excluded for ASASSN-18bt, a majority of proposed \snia\ progenitor systems are still allowed. For comparison, mass loss rates of Galactic symbiotic systems, for an assumed wind velocity of 10 km s$^{-1}$, are shown as grey stars \citep{seaquist1990}.  Figure is adapted from \citet{margutti14}.}
\label{fig:Xray}
\end{figure}

In Figure~\ref{fig:Xray}, we compare this limit to other constraints on the density surrounding nearby SN~Ia from X-ray observations \citep{margutti12, russell12, margutti14, shappee14} as well as the expectations for a variety of proposed SN~Ia progenitor systems. Our limit is consistent with those found by \citet{russell12} for a large sample of SN~Ia observed with \swift/XRT, but approximately 3-4 orders of magnitude less constraining than the deep limits obtained from \emph{Chandra} observations of the nearby SN\,2011fe \citep{margutti12} and SN\,2014J \citep{margutti14}. As a result, while the \swift/XRT limit rules out a fraction of symbiotic progenitor systems for ASASSN-18bt, we do not expect to detect signatures from the range of main sequence and subgiant companions allowed by the early {\em Kepler}  light curve (Section~\ref{sec:companions}).

\section{Conclusions}
\label{sec:summary}

ASASSN-18bt is the nearest and brightest supernova detected by {\em Kepler}  to date yielding a light curve with a cadence and photometric precision better than that for any other SN~Ia light curve.  Our fit to the very early portion of the light curve unambiguously shows a nearly linear phase, a kink, and then a steeper rise that cannot be well-fit by a single power-law model.  An empirical double-power-law model fits the data reasonably well, hinting that two physical processes must be responsible for the observed rise.  Thus, ASASSN-18bt joins a growing list of SNe Ia whose early light curves are not well described by a single power law (e.g., SN~2012fr \citep{contreras18}, SN~2013dy \citep{zheng13}, SN~2014J \citep{goobar15, siverd15}), MUSSES1604D \citep{jiang17}, iPTF16abc \citep{miller18}, and DLT~17u \citep{hosseinzadeh17}).   This may be a common feature of SNe Ia that was not previously seen because high-cadence early observations of bright SNe have only become possible with the recent proliferation of high-cadence transient surveys like ASAS-SN, ATLAS, PTF, LOSS, and DLT40.  

We compared the ASASSN-18bt light curves to theoretical models of three physical processes that could affect the rising light curve of a SNe Ia. 

1) We first compared the early-time light curve to the companion interaction models of \citet{kasen10} for companions of various radii.  We found that a single power-law rise with a companion of any radius cannot reproduce the observed {\em K2}  light curve of ASASSN-18bt (Figure~\ref{fig:companion}).  We then simultaneously fit a double power law with a companion model and found nearly identically good fits from 0.01-8 \rsun{} companions assuming a favorable viewing angle.  This is because the first power law and the companion model can compensate for each other and that the dark time, the power-law index, and the companion radius are degenerate.  Thus, with fine tuning it is possible for the power law to conspire to hide the shock signature in a smooth curve.  This weak constraint on the progenitor system demonstrates that a better, physically motivated model for the rising SN light curve is required before we can confidently and robustly use early time light curves of SNe Ia to constrain their progenitor systems. 

2) We also compared the early light curve of ASASSN-18bt to models assuming different amounts of $^{56}$Ni mixing \citep{piro16,contreras18}.   The amount of mixing affects the diffusion time for energy released by radioactive decay and thus the early rise of the light curve. We find that at times less than 3 days after explosion, the light curve fits highly mixed $^{56}$Ni models, with a $^{56}$Ni mass fractions of $0.15-0.2$ at approximately 0.05~\msun\ below the surface of the progenitor WD, and at later times it is more consistent with a moderately mixed model. No single smooth $^{56}$Ni distribution accounts for the early light curve, though a non-smooth distribution may be able to do so.  
We then compared ASASSN-18bt to the synthetic light curves from \citet{noebauer17} for a variety of explosion models. We found that only the double-detonation model, with its small amount of surface radioactive material, can qualitatively match the rise for the first few days. We note, however, that other models not tested in this work (e.g., collision models; \citealp{dong15,dong18}) may also produce similar features in the early-time light curves if they produce small amounts of shallow $^{56}$Ni.

However, the effect that $^{56}$Ni in the outer ejecta has on other observations, like the spectroscopic evolution near maximum light, must carefully be considered (e.g., \citealp{nugent97, kromer10, woosley11}).  Perhaps the most direct observation evidence for this material is the claimed detection of the 158 keV $^{56}$Ni gamma-ray decay lines between 16-35 days after explosion in the nearby SN~2014J \citep{diehl14, isern16}.  At these phases the ejecta is expected to be optically thick at these wavelengths and therefore emission from this line is expected from radioactive material located in the very outer layers.  Current work in the literature suggest the measured line flux requires $\sim 0.06$ \citep{diehl14} to $\sim 0.03-0.08$ \msun{} \citep{isern16} of $^{56}$Ni in the outer eject. Furthermore, similar to ASASSN-18bt, the rise of SN~2014J cannot be explained by a single power law \citep{goobar15, siverd15}.

3) The interaction between supernova ejecta and nearby CSM will also affect the early light curve of a SNe Ia. Even though nearly arbitrarily complex light curves are possible with complex distributions of nearby material,  \citet{piro16} argue that nearby circumstellar material will likely be distributed as $\rho \propto r^{-3}$.  We compared the light curve of ASASSN-18bt to the theoretical light curves presented in \citet{piro16} and find that none adequately reproduce the initial $\sim4$ day nearly linear rise observed in ASASSN-18bt.  However, more detailed theoretical studies are needed to fully explore the range of light curves that are feasible for physically motivated distributions of CSM material.

The absence of X-ray emission from ASASSN-18bt in \swift{} X-ray observations constrains the circumstellar material at much large distances and lower densities.  For a constant density CSM X-ray limits constrain $\rho_{\textrm{CSM}}<\rm{4.5 \times 10^{5}\,\rm{cm^{-3}}}$ at a radius of 4 $\times$ 10$^{15}$ cm and a progenitor wind to have $\dot M<\rm{8 \times 10^{-6}}\,\rm{M_{\sun}yr^{-1}}$ for $v_\textrm{w}=100\,\rm{km\,s^{-1}}$, at a radius of 4.5 $\times$ 10$^{15}$ cm from the progenitor star.  While the \swift/XRT limit rules out a fraction of symbiotic progenitor systems for ASASSN-18bt, the X-ray observation were not sensitive enough to detect accretion winds from main sequence and subgiant companions.

The early time light curves of SNe~Ia may finally help resolve the uncertainty of the progenitor systems of these prolific, energetic explosive events. There is a growing class of SNe~Ia with linearly rising early-time light curves for the first couple days which then steepen.  The cause of this feature is still unclear.  Without the well-sampled {\em K2}  light curve presented in this work for ASASSN-18bt the physical nature of this signature could have been confused or mis-interpreted. This discovery highlights the need for more theoretical work on the expected signatures from various progenitor models.  Additionally, significantly more observational work is needed to find nearby SNe Ia within the first $\sim$day of $t_{\textrm{1}}$ when interesting physical effects are not yet swamped by the $^{56}$Ni-power rising light curve.  However, this work also highlights the power of well-sampled early-time data and that immediate, multi-band, high-cadence followup will be needed for progress in our understanding SNe~Ia to continue.  With the recently expanded, now operational, next generation of public, all-sky transient surveys, having increased cadence and sensitivity (listed in Table~\ref{tab:surveys}), the collection of well-sampled light curves is expected to explode. Indeed, at the writing of this manuscript two SNe~Ia have already been discovered in the TESS field-of-view (ASASSN-18rn and ASASSN-18tb) where similar studies to this work will be performed.

\begin{deluxetable}{llcrr}
\tablewidth{240pt}
\tabletypesize{\footnotesize}
\tablecaption{All-sky Public Surveys}
\tablehead{
\colhead{Survey} &
\colhead{Hemispheres} &
\colhead{Number of} &
\colhead{Depth} & 
\colhead{Cadence} \\ 
\colhead{} &
\colhead{} &
\colhead{Sites}  &
\colhead{mag} &
\colhead{hours} }
\startdata
ASAS-SN    & N+S  & 4   & $\sim 18.5$   & 20   \\
ATLAS      & N    & 2   & $\sim 19.5$   & 48   \\
ZTF        & N    & 1   & $\sim 20.5$   & 72   \\
Pan-STARRS & N    & 1   & $\sim 22.0$   & 240  
\enddata \tablecomments{Rough survey parameters for the recently expanded real-time, all-sky surveys announcing discoveries to the community.  One can easily see how each survey complements the others in terms of cadence and depth.  Pan-STARRS cadence was estimated from best-case in \citet{weryk16}.} 
\label{tab:surveys} 
\end{deluxetable}

\acknowledgments

We thank Mark Phillips and Tony Piro for fruitful discussions and J.~C.~Wheeler and S.~J.~Smartt for their comments on the manuscript. Additionally, we thank the referee for their careful comments that have undoubtedly improved this work.   MD is supported by NASA through Hubble Fellowship grant HF-51348.001 awarded by the Space Telescope Science Institute, which is operated by the Association of Universities for Research in Astronomy, Inc., for NASA, under contract NAS 5-26555. MDS is supported by a research grant (13261) from VILLUM FONDEN.  CSK and KZS are supported by NSF grants AST-1515876 and AST-1515927. SD acknowledges Project 11573003 supported by NSFC. Support for JLP is provided in part by the Ministry of Economy, Development, and Tourism's Millennium Science Initiative through grant IC120009, awarded to The Millennium Institute of Astrophysics, MAS. TAT is supported in part by Scialog Scholar grant 24215 from the Research Corporation. EB and JD were supported in part by NASA grant NNX16AB25G. Work by S.V.Jr. is supported by the David G. Price Fellowship for Astronomical Instrumentation and by the National Science Foundation Graduate Research Fellowship under Grant No. DGE-1343012. Parts of this research were supported by the Australian Research Council Centre of Excellence for All Sky Astrophysics in 3 Dimensions (ASTRO 3D), through project number CE170100013.  This research was made possible through the use of the AAVSO Photometric All-Sky Survey (APASS), funded by the Robert Martin Ayers Sciences Fund.

We thank the Las Cumbres Observatory and its staff for its continuing support of the ASAS-SN project.  ASAS-SN is supported by the Gordon and Betty Moore Foundation through grant GBMF5490 to the Ohio State University and NSF grant AST-1515927. Development of ASAS-SN has been supported by NSF grant AST-0908816, the Mt. Cuba Astronomical Foundation, the Center for Cosmology and AstroParticle Physics at the Ohio State University, the Chinese Academy of Sciences South America Center for Astronomy (CASSACA), the Villum Foundation, and George Skestos.

This research has made use of the NASA/IPAC Extragalactic Database (NED) which is operated by the Jet Propulsion Laboratory, California Institute of Technology, under contract with the National Aeronautics and Space Administration.  This research has made use of NASA's Astrophysics Data System Bibliographic Services.  IRAF is distributed by the National Optical Astronomy Observatory, which is operated by the Association of Universities for Research in Astronomy (AURA) under a cooperative agreement with the National Science Foundation.  


\bibliographystyle{apj}

\end{document}